\documentclass[rmp,aps,reprint,longbibliography]{revtex4-2}
\setcitestyle{numbers,square,comma}

\usepackage{babel,calc,amsmath,amsthm,amssymb,mathdots,graphicx,subfigure}
\usepackage[table]{xcolor}
\usepackage[T1]{fontenc}
\usepackage[unicode=true]{hyperref}
\usepackage{mathtools}
\usepackage{booktabs}
\definecolor{ustcblue}{rgb}{0.0117, 0.3059, 0.6314}
\hypersetup{
     colorlinks=true,       		
     linkcolor=ustcblue,          	
     citecolor=ustcblue,             
     urlcolor=ustcblue,          
 }
 
\newcommand{\abs}[1]{\left\vert#1\right\vert}
\newcommand{\ket}[1]{\left\vert#1\right\rangle}
\newcommand{\bra}[1]{\left\langle#1\right\vert}
\newcommand{\braket}[1]{\left\langle#1\right\rangle}

\newcommand{\Tr}{{\rm tr}}

\makeatletter
\def\ignorecitefornumbering#1{%
     \begingroup
         \@fileswfalse
         #1
         \fi
    \endgroup
}
\makeatother

\newif\ifdebug

\ifdebug
\usepackage[normalem]{ulem}
\definecolor{zhliu}{rgb}{0.12, 0.72, 0.36}
\definecolor{marker}{rgb}{0.96, 0.48, 0}
\newcommand{\add}[1]{\textcolor{zhliu}{#1}}
\newcommand{\note}[1]{\textcolor{marker}{#1}}
\newcommand\delete{\bgroup\markoverwith{\textcolor{marker}{\rule[0.5ex]{2pt}{0.8pt}}}\ULon}

\else
\newcommand{\add}[1]{#1}
\newcommand{\note}[1]{\ignorespaces}
\newcommand{\delete}[1]{\ignorespaces}

\fi

\begin{document}
\renewcommand{\figurename}{Fig.}
\renewcommand{\tablename}{Table}
\renewcommand{\figureautorefname}{Fig.\!}
\renewcommand{\sectionautorefname}{Sec.\!}
\def\equationautorefname~#1\null{Eq.\,(#1)\null}

\title{Twenty years of quantum contextuality at USTC}

\author{Zheng-Hao~Liu}

\author{Qiang~Li}

\author{Bi-Heng~Liu}

\author{Yun-Feng~Huang}

\author{Jin-Shi~Xu}
\email{jsxu@ustc.edu.cn}

\author{Chuan-Feng~Li}
\email{cfli@ustc.edu.cn}

\author{Guang-Can~Guo}
\affiliation{CAS Key Laboratory of Quantum Information, University of Science and Technology of China, Hefei 230026, People's Republic of China}
\affiliation{CAS Centre for Excellence in Quantum Information and Quantum Physics, University of Science and Technology of China, Hefei 230026, People's Republic of China}
\affiliation{Hefei National Laboratory, University of Science and Technology of China, Hefei 230088, People's Republic of China}

\date{\today}

\begin{abstract}
Quantum contextuality is one of the most perplexing and peculiar features of quantum mechanics. Concisely, it refers to the observation that the result of a single measurement in quantum mechanics depends on the set of joint measurements actually performed. The study of contextuality has a long history at University of Science and Technology of China (USTC). Here we review the theoretical and experimental advances in this direction achieved at USTC over the last 20 years. We start by introducing the renowned simplest proof of state-independent contextuality. We then present several experimental tests of quantum versus noncontextual theories with photons. Finally, we discuss the investigation on the role of contextuality in general quantum information science and its application in quantum computation.
\end{abstract}

\maketitle

\tableofcontents

\section{Introduction}
\label{sec:introduction}

In its less than one hundred years of history, quantum mechanics has greatly changed the human society. The marriage of quantum mechanics and information theory have give birth to the novel interdisciplinary research field of quantum information science. Thanks to the ongoing ``second quantum revolution'' which has further improved our ability to operate single quantum entities, quantum technology can find its rule in multiple aspects of contemporary science. Especially, quantum computation holds the promise of enormous advancements in human's computational power \cite{Shor94}, and the state-of-the-art quantum computers have exhibited decisive speed-up \cite{hszhong20,Xanadu21,hszhong21} on specific tasks comparing with classical computers.

An intriguing observation about quantum computation is that while some behaviors of quantum circuit are particularly hard for supercomputers to reproduce \cite{Aaronson11,Hamilton17}, some features are classically simulable \cite{Gottesman98,Braunstein05} and thus do not provide a quantum speedup. The separation here can be traced back to a counter-intuitive phenomenon in the quantum foundation: contextuality (throughout this paper, we shall omit the qualifier ``quantum'' before ``contextuality'' for brevity). Contextuality has become a central concept in modern quantum information science: Not only does it engender a lot of quantum paradoxes \cite{GHZ89,Hardy92,Leifer05,sxyu14,Abramsky15,Waegell18,zhliu20cat}, it also serves a resource for many quantum information processing tasks. In particular, research in recent years has unraveled the connection between contextuality and universal quantum computation \cite{Howard14,Spekkens08,Raussendorf13}. In this setting, the study on contextuality helps both the comprehension of the quantum foundations and the further development of future quantum information technology. 

Comparing with its broad applications and clear significance, the concept of contextuality itself is rather abstract and comes with heavy mathematical background. Historically, the discovery of contextuality is inspired by the debate of the completeness of quantum theory \cite{EPR35} and the difficulty of reformulating it with hidden-variable theories. \citet{KS67} first established that such a difficulty lies in that a hidden-variable description of quantum measurements must be context-sensitive, so it is not possible to reconcile quantum theory with noncontextual hidden-variable models. The result is now known to the quantum community as the Kochen--Specker theorem. Among many theoretical topics in the study of contextuality, a renowned question is to simplify the proof of the Kochen--Specker theorem, so it utilizes less measurements, becomes more robust against noise, and does not rely on the selected quantum state to manifest contextuality. To date, this theoretical investigation has seen fruitful outcome: \citet{YO12} have proposed an elegant state-independent proof of contextuality, which makes use of provably the fewest measurements \cite{Cabello15}. 

The other complementary approach to the study of contextuality is to design and implement experiments to directly test the conflict between quantum and classical theories. However, to this purpose, it is not enough to measure the overall probability distribution of a quantum system under several measurement bases---consecutive measurements are almost always necessary to track down the evolution of a single quantum system, posing high requirements on contextuality experiments. The work by \citet{yfhuang03} is among the earliest experimental tests of contextuality. Since then, researchers have striven to carry out experimental works on various forms of contextuality and investigate its implications in the broader area of quantum information science. Among these works, many chose the linear optics platform \cite{Flamini18} as the experimental system due to its capability of making high-precision quantum state preparation, transformation and measurement, its long coherence time and its rich intrinsic degrees of freedom facilitating complicated forms of quantum operations---all indispensable resource for the experimental study of contextuality.

Over the past twenty years, USTC have overseen the rapid development of the study of contextuality. The central role of contextuality in quantum information science, and the plentiful theoretical and experimental results achieved here are the dual motivations of this review. Because contextuality is a broad topic with a myriad of results and perspectives, it is impossible for us authors to present all the exciting results here. For the enthusiastic readers, we point to \cite{Budroni21} for a more comprehensive review of contextuality, and \cite{Thompson16} for advances in contextuality tests. The remainder of this paper is arranged as follows. In \autoref{sec:theory}, we review the elegant proof of contextuality by \citet{YO12}, and accompany this result with the recently developed, systematic method of discovering new proofs of contextuality from a graph-theoretical approach \cite{CSW14} and give an example. In \autoref{sec:experiment}, we present several representative experimental tests of contextuality by USTC groups on the linear optics platform, demonstrating the suitability of the photonic architecture as a testbed of quantum foundations. In \autoref{sec:application}, we discuss the role of contextuality in quantum foundation, quantum information science and quantum computation based on further experimental works. Finally, in \autoref{sec:discussion}, we give a brief summary of the results and envisage the potential development in this vibrant research field.

\section{Theory: the Kochen--Specker theorem and its proofs}
\label{sec:theory}

We begin by introducing the Kochen--Specker theorem. It reveals the impossibility of describing a quantum measurement in a Hilbert space dimension $d\geqslant3$ using a noncontextual hidden-variable model. Here, the term ``noncontextual'' indicates that the model relies only on the physical state, possibly plus a hidden variable that can be absorbed in the state, and the projector defining the projective measurement itself, instead of the entire set of projectors that forms an orthogonal measurement, that we denote as the  ``context''. \add{We note that here only contextuality with rank-1 projectors are considered; the recently developed generalized contextuality based on positive operator-valued measurements \cite{zpxu21rank} are not discussed in this review.}

It is beneficial to express the Kochen--Specker theorem using the terminologies in quantum measurements. An orthogonal measurement is composed of a series of orthogonal projectors $\hat{\mathbf{\Pi}}=\{\hat{\Pi}_1, \hat{\Pi}_2, \ldots, \hat{\Pi}_d\}$, with $\hat{\Pi}_i\hat{\Pi}_j=\hat{\Pi}_i\delta_{ij}$ and $\sum_{k=1}^d \hat{\Pi}_k = \mathbb{I}_d$, where $d$ is the dimension of the Hilbert space spanned by these projectors. When an orthogonal measurement is cast on a quantum state $\rho$, it evolves the state to the nondegenerate eigenstate of a random projector $\hat{\Pi}_k\in\hat{\mathbf{\Pi}}$ according to the L\"uders' rule: 
\add{
\begin{align}
    \rho\to\rho^\prime=\displaystyle\sum_{k=1}^d\hat{\Pi}_k \rho\hat{\Pi}_k,
\end{align} 
}with a probability specified by the Born's rule: \begin{align}
    \Pr(k)=\Tr(\rho\hat{\Pi}_k).
\end{align}
We see the randomness in measurement is an intrinsic feature of quantum theory. The measurement outcome can still be indeterministic even for a pure quantum state provided that $\rho\hat{\mathbf{\Pi}}$ has nonunit rank.

\begin{table*}[tbh]
    \centering
    \begin{tabular}{ccccccccc}
    \toprule\toprule
        $C_1$ & $C_2$ & $C_3$ & $C_4$ & $C_5$ & $C_6$ & $C_7$ & $C_8$ & $C_9$ \\
    \midrule
        $(1,0,0,0)$ & $(1,0,0,0)$ & $(0,1,\bar{1},0)$ & $(\bar{1},1,1,1)$ & $(0,1,0,\bar{1})$ & $(0,0,1,\bar{1})$ & $(0,0,1,1)$ & $(0,1,0,0)$ & $(0,1,1,0)$ \\
        $(0,0,1,\bar{1})$ & $(0,1,\bar{1},0)$ & $(\bar{1},1,1,1)$ & $(0,1,0,\bar{1})$ & $(1,1,1,1)$ & $(1,1,1,1)$ & $(1,1,1,\bar{1})$ & $(0,0,0,1)$ & $(1,0,0,1)$ \\
        $(0,0,1,1)$ & $(0,0,0,1)$ & $(1,1,1,\bar{1})$ & $(1,1,\bar{1},1)$ & $(1,0,\bar{1},0)$ & $(1,\bar{1},0,0)$ & $(1,1,\bar{1},1)$ & $(1,0,1,0)$ & $(1,\bar{1},1,\bar{1})$ \\
        $(0,1,0,0)$ & $(0,1,1,0)$ & $(1,0,0,1)$ & $(1,0,1,0)$ & $(1,\bar{1},1,\bar{1})$ & $(1,1,\bar{1},\bar{1})$ & $(1,\bar{1},0,0)$ & $(1,0,\bar{1},0)$ & $(1,1,\bar{1},\bar{1})$ \\
    \bottomrule\bottomrule
    \end{tabular}
    \caption{A Kochen--Specker set of rays by \ignorecitefornumbering{\citet{Cabello96}}. The normalization of the rays are omitted and $\bar{1}$ denotes $-1$. \add{For example, if the bases are chosen as $\{\ket{i_1}, \ket{i_2}, \ket{i_3}, \ket{i_4}\}$, then the ray $(0,0,1,\bar{1})$ corresponds to a state vector $(\ket{i_3}-\ket{i_4})/\sqrt{2}$.} These rays as directions of projective measurements can reveal the conflict between quantum and noncontextual theories: if the results of the measurements are predetermined before the measurements actually take place, the outcomes ``0'' and ``1'' corresponding to a single run of experiment can be assigned to each rays, and the assignment throughout the table should be consistent \cite{Fine82}. Note that every ray appears twice in the table, so in total an even number of rays must be assigned ``1''; however, every column of the table forms an orthonormal basis, so the number of rays assigned ``1'' must be 9, contradiction.}
    \label{tab:CEG18}
\end{table*}

On the other hand, in a noncontextual hidden-variable theory the randomness of a quantum measurement can be attributed to the ignorance of the ontic state $\lambda$, that we call the hidden variable. Within this framework, the density matrix is a function of the ontic state: $\rho=\rho(\lambda)$, and the outcome of a projective measurement $\hat{\Pi}_k$ can be instead specified by a binary response function: 
\begin{align}
    v(\hat{\Pi}_k, \lambda)\in\{0,1\}, \;\text{with}\;\int v(\hat{\Pi}_k, \lambda) d\lambda=\Pr(k).
\end{align}In order to recover the orthogonality between quantum measurements, the supports of the response function for orthogonal projectors must be disjoint: 
\begin{align}
    v(\hat{\Pi}_i, \lambda)v(\hat{\Pi}_j, \lambda)\equiv0, \;\;\forall\;\hat{\Pi}_i\hat{\Pi}_j=0, \lambda.
    \label{eq:resp_ortho}
\end{align}
Also, the completeness of quantum measurement requires that if a response function is consistent with quantum predictions, it must satisfy:
\begin{align}
    \displaystyle\sum_{k=1}^d v(\hat{\Pi}_k, \lambda)\equiv1, \;\;\forall\;\displaystyle\sum_k\hat{\Pi}_k=\mathbb{I}, \lambda.
    \label{eq:resp_comp}
\end{align}
It is clear from the definition that the response function does not rely on the entire measurement context, and the hidden-variable description can recover the marginal distribution of every projective measurement. 

How does the difference between the quantum and hidden-variable theories manifest as an experimentally testable object, instead of staying at a metaphysical level? Many discussions have been devoted to address this question. According to the definition of the response function, it is context-insensitive and its value for a specific projector must be consistent for different choices of orthogonal measurements. By Fine's theorem \cite{Fine82}, the response function can then be extended to noncommutative projectors $[\hat{\Pi}_i, \hat{\Pi}_j]\neq0$. Therefore, in a noncontextual hidden-variable model the definition of response function is global: every projective measurement, commuting or not, can be assigned to a definite outcome prior to experiment. The difference between the quantum and hidden-variable theories can then be revealed by showing that such a global response function cannot always preserve the completeness and orthogonality of quantum measurement.

We quote the delicate construction by \citet{Cabello96} to illustrate the impossibility of such a definite value assignment in a real experiment. Consider the rays $\ket{r}$ in \autoref{tab:CEG18} which defines the set of projectors $\hat{\Pi}_r=\ket{r}\bra{r}$. We have the following observations:
\begin{enumerate}
    \item The four rays in each of the nine columns spans an orthonormal basis.
    \item Each ray appears twice in the entire table.
\end{enumerate}
The first observation indicates that, in every run of experiment, one and only one of the four response functions corresponding to the four rays in the same column returns 1. Therefore, across the entire \autoref{tab:CEG18}, nine rays will be assigned 1. However, the second observation indicates that, if the response function is consistent and context-insensitive, the number of rays assigned to 1 must be even. The contradiction between the two observations shows the impossibility of defining a global response function for the measurement outcomes of the projectors in \autoref{tab:CEG18}, and demonstrated the incompatibility between the quantum and hidden-variable theories' description on measurements.

\subsection{Towards the ultimate simplicity}

The simplification of the Kochen--Specker theorem's proofs appropriately reflects the development of the study on contextuality. When \citet{KS67} found the first proof of contextuality, the number of rays used in the proof was 117, making its comprehension extremely hard at that time. After almost thirty years' search, the number was finally able to be reduced to 18 \cite{Cabello96}, and it can be shown that the number of rays required cannot be further reduced, if the proof is based on the impossibility of defining the response function \cite{zpxu20peres}. Note however if we wish to demonstrate contextuality in three-dimensional Hilbert space, the number of required rays will be no less than 22 \cite{Uijlen16}, and the well-known construction by \citet{Peres91} utilizes 33 rays.

\begin{figure*}[tbh]
    \centering
    \includegraphics[width=.8\textwidth]{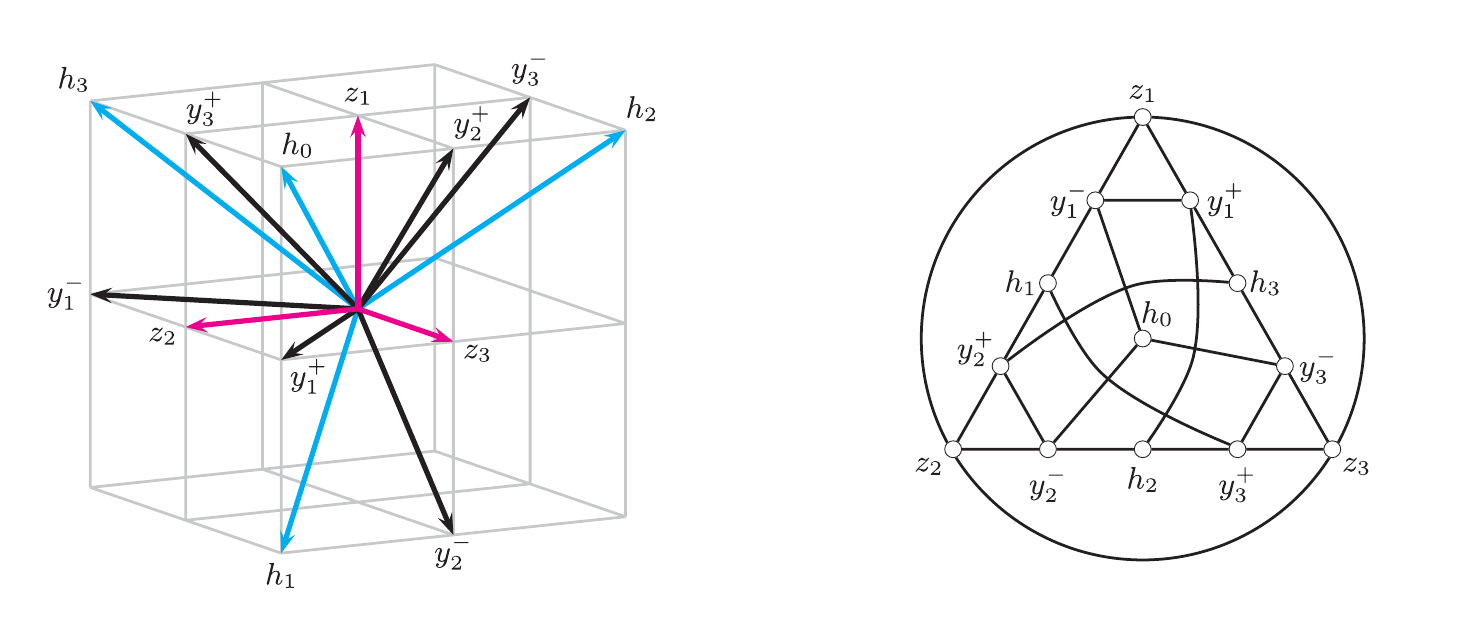}
    \caption{The Yu--Oh 13-ray appearing in the state-independent proof of contextuality by \citet{YO12}. Left: the geometric representation of the rays in a unit cube. The rays are defined as: $y_{1} ^{-} =(0,1,-1),\; y_{2} ^{-} =(-1,0,1),\; y_{3} ^{-} =(1,-1,0),\; y_{1} ^{+} =(0,1,1),\; y_{2} ^{+} =(1,0,1),\; y_{3} ^{+} =(1,1,0),\; h_{0} =(1,1,1),\; h_{1} =(-1,1,1),\; h_{2} =(1,-1,1),\; h_{3} =(1,1,-1),\; z_{1} =(1,0,0),\; z_{2} =(0,1,0),\; z_{3} =(0,0,1).$ Right: the orthogonality relationship among the set of rays. Each vertex represents to a ray; when two vertices are linked by an edge, the corresponding rays are orthogonal. Figure taken from Reference \ignorecitefornumbering{\cite{YO12}}.}
    \label{fig:yo_vec}
\end{figure*}

Is it possible to derive further simplified proofs of contextuality? In 2012, \citet{YO12} published a construction of state-independent contextuality using a set of only 13 rays and works for a three-dimensional quantum system. Subsequently, the construction was found to be optimal \cite{Cabello15} in the sense that the number of rays in a state-independent proof of contextuality cannot be further reduced. The set of rays, now renowned as the Yu--Oh 13-ray, are chosen from the vertices, face centers and edge centers of a cube.

The definition of the rays is shown in \autoref{fig:yo_vec}, together with a graph depicting the orthogonal relationships between these rays, where the vertices stand for the rays and the edges signify the orthogonal relationships between the rays corresponding to connected vertices. Using the graph representation, the restriction on the response functions, \autoref{eq:resp_ortho} and \autoref{eq:resp_comp} can be instead interpreted as a coloring rule on the vertices: (1) at most one in a pair of connected vertices is colored (assigned ``1''); and (2) one and only one vertex in a $d$-clique is colored. If the graph representation can be properly colored, then a global response function will exist. For the representation of the Yu--Oh 13-ray, one can easily check out that such a coloring scheme exist, so the methodology in previous proofs of contextuality---by demonstrating the impossibility of defining the response function---does not apply here.

However, here lays the essence of Yu and Oh's result: a proof of contextuality can be derived even a global response function exists. This new framework of finding state-independent contextuality goes beyond the previous paradigm of searching for a Kochen--Specker set of rays; here, we shall paraphrase their reasoning to show how this is accomplished. Firstly, observe that if we apply the color rule described in the last paragraph, only one of the four vertices $h_k,\;k\in\{0,1,2,3\}$ can be colored. This can be explicate by the following two \textit{reductio ad absurdum} arguments: 
\begin{enumerate}
    \item Suppose $h_1$ and $h_2$ are both colored, then $y_1^\pm$ and $y_2^\pm$ must be uncolored. By completeness, $z_1$ and $z_2$ must both be colored, but they are connected by edges and thus cannot be both colored, contradiction.
    \item Suppose $h_0$ and $h_1$ are both colored, then $y_2^\pm$ and $y_3^\pm$ must be uncolored. By completeness, $z_2$ and $z_3$ must both be colored, but they are also connected by edges so, again, contradiction.
\end{enumerate}
By the $C_3$-symmetry of the graph, the arguments apply on any choices of vertices $h_k$. Consequently, the response function of the four projective measurements must have no common support, and the total probability of finding a quantum state on the four projectors cannot be more than 1. Secondly, consider the projectors corresponding to the rays $h_k$ in \autoref{fig:yo_vec} and take the sum of these projectors yields: 
\begin{align}
    \ket{h_0}\bra{h_0}+\ket{h_1}\bra{h_1}+\ket{h_2}\bra{h_2}+\ket{h_3}\bra{h_3}=\frac{4}{3}\mathbb{I}_3,
\end{align}
that is, the total probability of finding an arbitrary quantum state on the four projectors is $4/3>1,$ in stark contrast to the predictions of the noncontextual hidden-variable theory. This completes the proof of contextuality with the Yu--Oh 13-ray.

Besides establishing the paradigm of state-independent contextuality beyond Kochen--Specker set, the Yu--Oh 13-ray also have other merits. Firstly, as already mentioned above, the proof only involving 13 rays sets the final record of minimal number of rays required to observe contextuality. Secondly, the proof works for a three-dimensional indivisible system, the smallest system showing contextuality, and thus have strong universality. Thirdly, the relation of orthogonality between the 13 rays actually implies a state-independent noncontextuality inequality (that will be discussed in \autoref{sec:experiment}) even without assuming \autoref{eq:resp_comp}, the requirement of completeness. The merit is the noncontextuality inequality is theory-independent and does not rely on any assumptions in quantum mechanics, in a similar vein with the inequalities in Reference \cite{Cabello08}. Fourthly, the derived inequality only involves marginal probabilities and two-point correlations between the projectors in \autoref{fig:yo_vec} and greatly facilitates its experimental test; in comparison, previously studied state-independent noncontextuality inequalities \cite{Cabello08} always require measurements of no less than three-point correlations. Finally, the methodology of analyzing the response function with the graph representation would soon exhibit its power and be developed into a more general framework, namely, the graph-theoretical approach to contextuality, which we shall discuss subsequently.

\subsection{The graph-theoretical approach to contextuality}

The graph-theoretical approach to contextuality, developed by \citet{CSW14} provides a generic method to construct noncontextual hidden-variable inequalities using the orthogonal relationships between rays corresponding to projective measurements. More specifically, it gives the describes how to compute the strongest correlations allowed by noncontextual hidden-variable and quantum theory with a given set of projection measurements.

We start by the formal definition of the \textit{graph of exclusivity} that is a central concept in this approach; the orthogonality graph in \autoref{fig:yo_vec} is already a graph of exclusivity. Given a set of abstract measurements $\tilde{\Pi} _k, \;k\in\{1, \ldots, n\}$, the graph of exclusivity corresponding to the set of measurements is an undirected graph $G=G(V, E)$, such that the vertex set $V(G)$ of the graph and the set of measures $\tilde{\bm\Pi} $ has a one-to-one correspondence, $|V(G)|=n$, and the edges of the graph connect vertices corresponding to mutually exclusive abstract measurements: $(i,j)\in E(G),\;\; \forall\; \tilde{\Pi} _i\tilde{\Pi} _j=0$. For quantum measurement, the abstract measurement operators are just the projectors: \begin{align}
    \tilde{\Pi} _k\to\hat{\Pi} _k=\ket{v_k} \bra{v_k}.
\end{align}
For noncontextual hidden-variable theory, although the form of measurement operators cannot be written explicitly at this time, the response functions corresponding to mutually exclusive measurement operators also satisfy a very simple relationship: 
\add{
\begin{align}
    v(\tilde {\Pi}_i , \lambda) + v(\tilde{\Pi}_j , \lambda) \leqslant 1,\;\; \forall\; \lambda,\; (i,j)\in E(G),
\end{align}
}suppose otherwise, then at least some $\lambda$ makes both measurements respond to 1, contradicting the requirement of orthogonality.

The graph-theoretical approach to contextuality links the maximally allowed quantum and noncontextual correlation to the constants of the graph of exclusivity. 
\add{
For the noncontextual correlations, it is:
\begin{align} 
\sum_{k=1}^n \Pr (\tilde{\Pi}_k) \overset{\rm NCHV} {\leqslant}  \alpha(G),
\end{align} 
with $\alpha(\cdot)$ being the independence number of a graph, defined as the cardinality of its largest subset of mutually disjoint vertices. For the quantum correlation, it is:
\begin{align} 
     \sum_{k=1}^n \Pr (\tilde{\Pi}_k) \overset{\rm Q} {\leqslant} \vartheta(G).
\end{align}  
Here, $\vartheta(\cdot)$ is the Lov\'asz number defined as: 
\begin{align}
    \vartheta(G):= \max_{\phi_k,\psi}  \sum_{k=1} ^n \left\vert \braket{\phi_k\vert\psi}  \right\vert^2,
\end{align}
note that the rays $\phi_k$ can be arbitrary chosen so long as the relations of exclusivity are satisfied. According to Lov\'asz's sandwich theorem \cite{Lovasz79}, it is not less than the independence number of the same graph: 
\begin{align}
\vartheta(G)\geqslant\alpha(G),\; \forall\, G.
\end{align}
}Therefore, the sum of probability allowed by a set of measurements with known orthogonality can be efficiently bounded with the graph-thoretic constants of the measurements' graph of exclusivity.
In addition, to test contextuality with realistic quantum measurements and incorporate the imperfection of orthogonality, the following noncontextuality inequality \cite{Cabello16} is shown to be tight:
\begin{align} 
    \sum_{k} \Pr (1|k) - \sum_{(i,j)\in E(G)} \Pr (1,1|i,j) \overset{\rm NCHV} {\leqslant}  \alpha(G),
    \label{eq:CSW_exp} 
\end{align}
here, $\Pr (1|k)=\Pr (\tilde{\Pi}_k)$ denotes the probability of the measurement $\tilde{\Pi}_k$ returning 1, and $\Pr (1,1|i,j)$ denotes the probability of the measurements on a pair of ideally exclusive measurements $\tilde{\Pi}_i$ and $\tilde{\Pi}_j$ both returning 1.

\begin{figure*}[tbh]
    \centering
    \includegraphics[width=.88\textwidth]{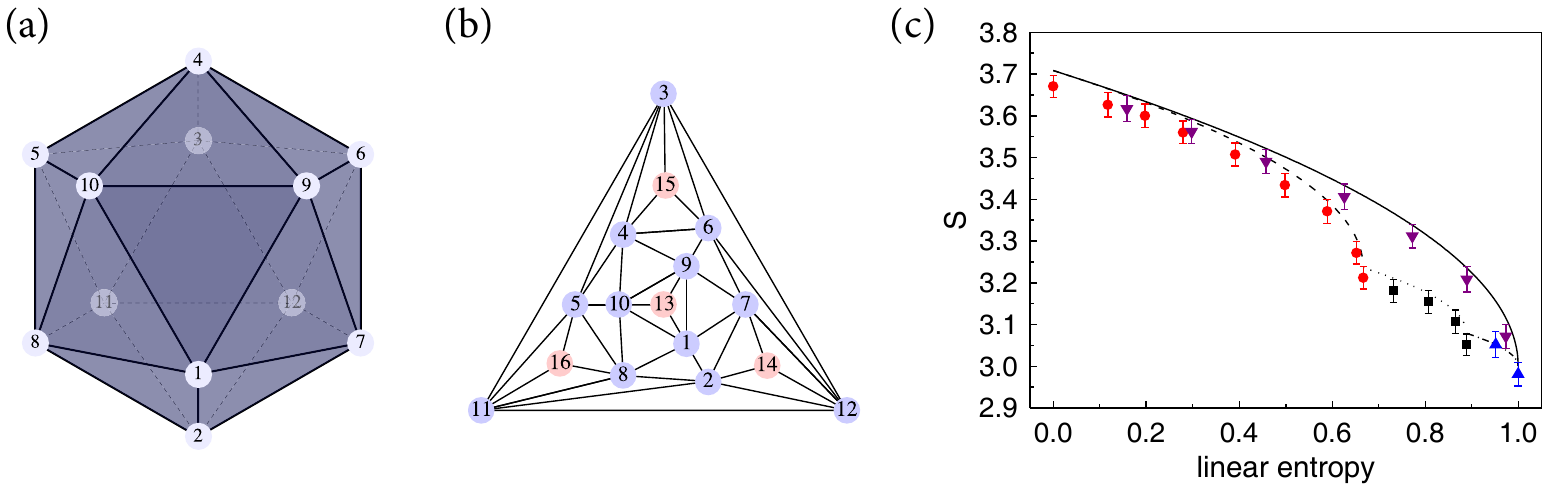}
    \caption{Contextuality from a Platonic graph. (a) A regular icosahedron is a Platonic solid with 12 vertices and 20 edges. (b) The icosahedron graph (vertices 1--12) is the skeleton of the icosahedron. With the auxiliary vertices 13--16 every vertex belongs to a 4-clique, and the graph's complement graph has a Lov\'asz orthogonal representation \cite{Lovasz89} in dimension 4. (c) The violation of the noncontextuality inequality dual to the icosahedron graph decreases with the linear entropy of a quantum state characterizing the mixedness of the state. Figure taken from Reference \ignorecitefornumbering{\cite{yxiao18platonic}}.}
    \label{fig:plato}
\end{figure*}

Several experimental works \cite{Canas16,yxiao17,zhliu19,yxiao18platonic} have followed this approach. For example, \citet{yxiao17} utilized the graph of exclusivity corresponding to the Yu--Oh 13-rays to derive and test an optimal state-independent noncontextuality inequality; \citet{zhliu19} adopted an exclusivity graph originated from a three-setting Bell inequality to demonstrate a contextual correlation stronger than Bell nonlocality. Because requirements other than orthogonality of measurements are not required in the graph-theoretical approach to contextuality, it can be used to discover previously unknown noncontextuality inequalities by devising exclusivity graphs with large quantum--classical ratio $\vartheta/\alpha$; the quantum states and measurements that maximally violate the inequalities can in turn be efficiently searched with semidefinite programming. 

Here, we explain this method in more detail by reviewing the work by \citet{yxiao18platonic}. The authors considered the Platonic graphs, the skeletons of the five Platonic solids, as the graph of exclusivity. Through an exhaustive search, it was found that the skeleton of the dodecahedron and the icosahedron (see \autoref{fig:plato}(a)) induce meaningful noncontextuality inequality, i.e., $\vartheta/\alpha>1$. Especially, the icosahedron graph $G_{\rm I}$ in \autoref{fig:plato}(b) has a independence number of $\alpha(G_{\rm I})=3$ and a Lov\'asz number of $\vartheta(G_{\rm I})=3(\sqrt{5}-1)$, resulting in a pronounced quantum--classical ratio of $\vartheta/\alpha=\sqrt{5}-1\approx1.236$ larger than that of the pentagram ($\vartheta/\alpha=\sqrt{5}/2\approx1.118$) which was previously widely exploited to test contextuality \cite{KCBS08,Lapkiewicz11,Ahrens13,Jerger16}. Using \autoref{eq:CSW_exp}, the noncontextuality inequality here can be explicitly expressed as: 
\begin{align}
    {\cal I}=\sum_{k} \Pr (1|k) - \sum_{(i,j)\in E(G_{\rm I})} \Pr (1,1|i,j) \overset{\rm NCHV} {\leqslant} 3,
\end{align}
the inequality can be violated up to ${\cal I}\overset{\rm Q}{\leqslant}3(\sqrt{5}-1)$ using quantum measurement. Through semidefinite programming, it was found the set of state and measurement rays saturating the quantum maximum can be embedded in a 4-dimensional Hilbert space. The large quantum--classical separation and the relatively low requirement on system dimension makes the icosahedron inequality an excellent candidate for experimental test.

The noncontextuality inequality associated with the icosahedron graph $G_{\rm I}$ has another merit that the inequality is pseudo state-independent: although a set of no more than 12 rays cannot comprise a state-independent proof of contextuality, this icosahedron inequality can be violated by all but the maximally mixed quantum state, provided a set of projective measurements are appropriately chosen according to the input quantum state. The pseudo state-independence stems from the spectra of the projectors that are chosen to saturate the quantum bound: the eigenvalues of the sum of the projectors are $\{3(\sqrt{5}-1), 5-\sqrt{5}\}$ with the first one being the Lov\'asz number $\vartheta(G_{\rm I})$ and the second one threefold degenerate. For the maximally mixed state $\mathbb{I}_4$ the inequality evaluates to ${\cal I}=[3(\sqrt{5}-1)+3(5-\sqrt{5})]/4=3$; for any other state we can choose the projectors so the eigenvector of the sum of the projectors will be aligned with the dominant eigenvector of the quantum state. By doing so, the inequality will be violated by any quantum state other than the maximally mixed state. Furthermore, if we choose the linear entropy to quantify the mixedness of the state, defined as $\ell=4(1-\Tr(\rho^2))/3$ for ququart states, then it can be shown that the quantum value of the icosahedron inequality is upper bounded by ${\cal I}\overset{\rm Q}{\leqslant}3+(3\sqrt{5}-6)\sqrt{1-\ell}.$ Therefore, the icosahedron inequality as a noncontextuality inequality can also be considered a proxy to estimate the purity of a quantum state.

\section{Experiment: photonic tests of contextuality}
\label{sec:experiment}

In this section, we proceed to review the recent progress of contextuality tests on the photonic platform. As we shall elucidate below, the experimental tests of contextuality also develops into two complementary approaches. The first category of experiments simplifies the requirements of contextuality tests by introducing and justifying some additional assumptions. At the price of decreased stringency, this approach facilitates the tests of a vast family of noncontextuality inequalities. In contrast, the second category of experiments aims to strictly follow the requirements from the theoretical models, and closes the experimental loopholes for some celebrated forms of contextuality.

To further discuss the two paradigms, it is best to start from the seminal experimental work \cite{yfhuang03} which, despite being qualitative, caught the essence of contextuality. By this example, even the readers not initially familiar with contextuality can quickly establish the basic concepts of contextuality experiments. The work was based on two simple observation about the maximally entangled qubit state, $\ket{\Phi^+}=(\ket{00}+\ket{11})/\sqrt{2}$ in quantum theory \cite{Simon00}: firstly, it is one of the Bell states, hence it is the common eigenstate of the following Pauli-product operators: $(\sigma_x^1\otimes\sigma_x^2)\ket{\Phi^+}=+\ket{\Phi^+}$, $(\sigma_z^1\otimes\sigma_z^2)\ket{\Phi^+}=+\ket{\Phi^+}$, and $(\sigma_y^1\otimes\sigma_y^2)\ket{\Phi^+}=-\ket{\Phi^+}$. Therefore, the following assertions hold:
\add{
\begin{align}
    \braket{\sigma_x^1\otimes\sigma_x^2}_{\Phi^+}=\braket{\sigma_z^1\otimes\sigma_z^2}_{\Phi^+}=+1.
    \label{eq:phip_1}
\end{align}
}Here, we have used the shorthand notation $\braket{\cdot}_\psi:=\braket{\psi|\cdot|\psi}$ to denote the expectation value of an operator for a specific quantum state. Secondly, as $\sigma_y^1\otimes\sigma_y^2=(\sigma_x^1\otimes\sigma_z^2)\cdot(\sigma_z^1\otimes\sigma_x^2)$, it immediately follows that 
\add{
\begin{align}
    \braket{(\sigma_x^1\otimes\sigma_z^2)\cdot(\sigma_z^1\otimes\sigma_x^2)}_{\Phi^+}=-1.
    \label{eq:phip_2}
\end{align}
}Equivalently, the measurement results for $\sigma_x^1\otimes\sigma_z^2$ and $\sigma_x^1\otimes\sigma_z^2$ should be different---one being $+1$ while the other being $-1$. However, the quantum predictions \autoref{eq:phip_1} and \autoref{eq:phip_2} already exclude a noncontextual hidden-variable description. Indeed, the response functions for the observables in \autoref{eq:phip_1} must satisfy $v(\sigma_x^1\otimes\sigma_x^2)=1$ and $v(\sigma_z^1\otimes\sigma_z^2)=1$ \add{(here we omit the choice of the ontic state $\lambda=\Phi^+$ for the response function.)}. Since each of the observables are defined over two qubits, it is also physically plausible to split the bipartite response functions into which of indivisible operators: $v(\sigma_j^1\otimes\sigma_k^2)=v(\sigma_j^1)v(\sigma_k^2),\; j,k\in\{x,y,z\}$. By doing this and again re-combine the operators, we arrive at $v(\sigma_x^1\otimes\sigma_z^2)v(\sigma_z^1\otimes\sigma_x^2)=1$, that is, the measurement results for $\sigma_x^1\otimes\sigma_z^2$ and $\sigma_x^1\otimes\sigma_z^2$ should be the same---simultaneously $+1$ or $-1$. Therefore, a noncontextual hidden-variable theory will give opposite prediction as \autoref{eq:phip_2} when the constraints in \autoref{eq:phip_1} held.

\begin{figure}[t!]
    \centering
    \includegraphics[width=.99\columnwidth]{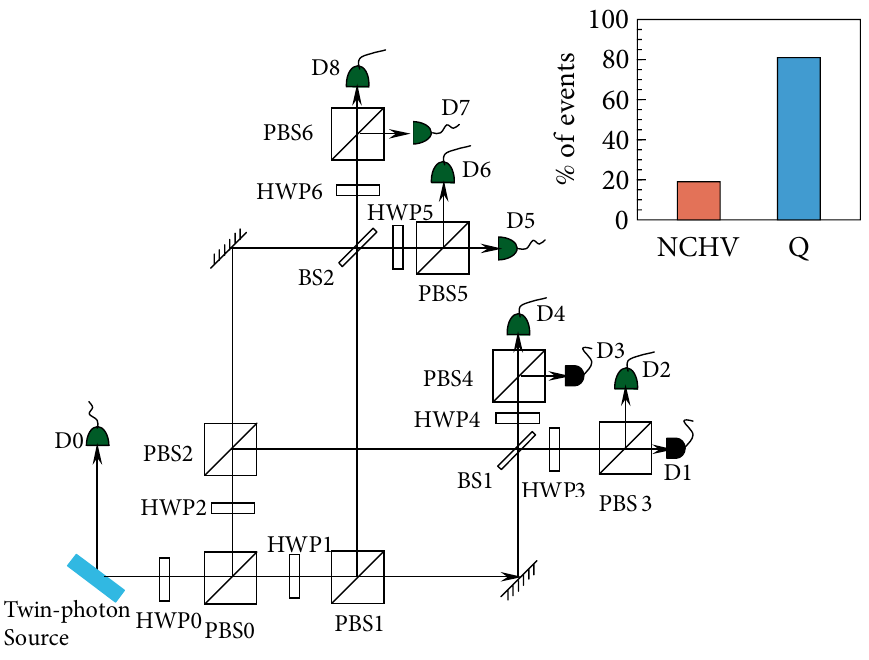}
    \caption{First experimental test of contextuality in USTC. Main: experimental setup. A heralded single photon's path and polarization degrees of freedom encoded two qubits. The half-wave plates and polarizing beam splitters inside the two Mach--Zehnder interferometers conducted the first joint path--polarization measurement, and that after the interferometers executed the second joint measurement. HWP half-wave plate and PBS polarizing beam splitter. Inset: experimental result showing event probabilities in accord with the predictions of the noncontextual hidden-variable and quantum theories. Figure adapted from Reference \cite{yfhuang03}.}
    \label{fig:exp_yfhuang03}
\end{figure}

In a pioneering work, \citet{yfhuang03} reported a direct experimental test of the above arguments. The experimental setup is shown in \autoref{fig:exp_yfhuang03}. The two-qubit state was encoded on the polarization and path states of a single photon: through a polarizing beam splitter (PBS0), the photons with vertical polarization state, $\ket{V}$, were reflected towards PBS2; we denote this path state as $\ket{R}$. The horizontally polarized $\ket{H}$ photons, on the other hand, still propagated towards PBS1; we label this path state as $\ket{L}$. Furthermore, the observables were defined as $\sigma_z^1=\ket{L}\bra{L}-\ket{R}\bra{R}, \sigma_z^2=\ket{H}\bra{H}-\ket{V}\bra{V}$, and the $\sigma_x$ operators accordingly. By adjusting the photon's initial polarization state with the half-wave plate HWP0, maximally entangled path--polarization state, $(\ket{HL}+\ket{VR})/\sqrt{2}$, was created.

For the determination of $\sigma_z^1\otimes\sigma_z^2$ and $\sigma_x^1\otimes\sigma_x^2$, the angle of the half-wave plates HWP1 and HWP2 were set as $0^\circ$. Due to the high extinction ratio of the polarizing beam splitters, $\sigma_z^1\otimes\sigma_z^2=+1$ was reasonably assumed. Subsequently, the measurement of $\sigma_x^1$ was implemented with a Mach--Zehnder interferometer between PBS0 and a balanced beam splitter BS1; the photons going toward PBS3 (PBS4) had $\sigma_x^1=+1(-1)$. Finally, at each output port of the interferometer, a half-wave plate set at $22.5^\circ$ assisted the polarizing beam splitter to realize the measurement of $\sigma_x^2$. For the determination of $\sigma_z^1\otimes\sigma_x^2$ and $\sigma_x^1\otimes\sigma_z^2$, the angle of HWP1 (HWP2) was changed to $22.5^\circ (-67.5^\circ)$ to introduce a Haramard operation on the polarization state to guarantee $\sigma_z^1\otimes\sigma_x^2=+1$. The PBS1 and PBS2 then implemented the measurement of $\sigma_z^2$. At the balanced beam splitters, the path information of the photon, $\sigma_x^1$, was transferred to the polarization degree of freedom, and was further read out again with the half-wave plates set at $22.5^\circ$ and the polarizing beam splitters.

Using the setup described above, the terms in \autoref{eq:phip_1} and \autoref{eq:phip_2} can be extracted according to a simple rule: in the ideal experimental setting, quantum theory predicts all photons to come to detectors labeled with odd numbers, while a noncontextual hidden-variable theory deems all photons to come to detectors labeled with even numbers. Therefore, the statistics of detector clicks directly tests the contextuality of quantum theory. Experimentally, it was found that 81\% of photons ended at odd-numbered ports; therefore, the result provided clear evidence for the contextual nature of quantum mechanics. 
From a modern perspective, we also notice that that the experiment suffered from several loopholes: the deduction of the first group of correlations required knowledge from quantum theory; the measurements of the same observable in different contexts utilized different apparatuses \cite{Amselem13}; most importantly, no testable noncontextuality inequality can be exploited to check the result and statistically refute the noncontextual models. These drawbacks will be solved in future works which are introduced in the following sections.

\begin{figure*}[tbh]
    \centering
    \includegraphics[width=.95\textwidth]{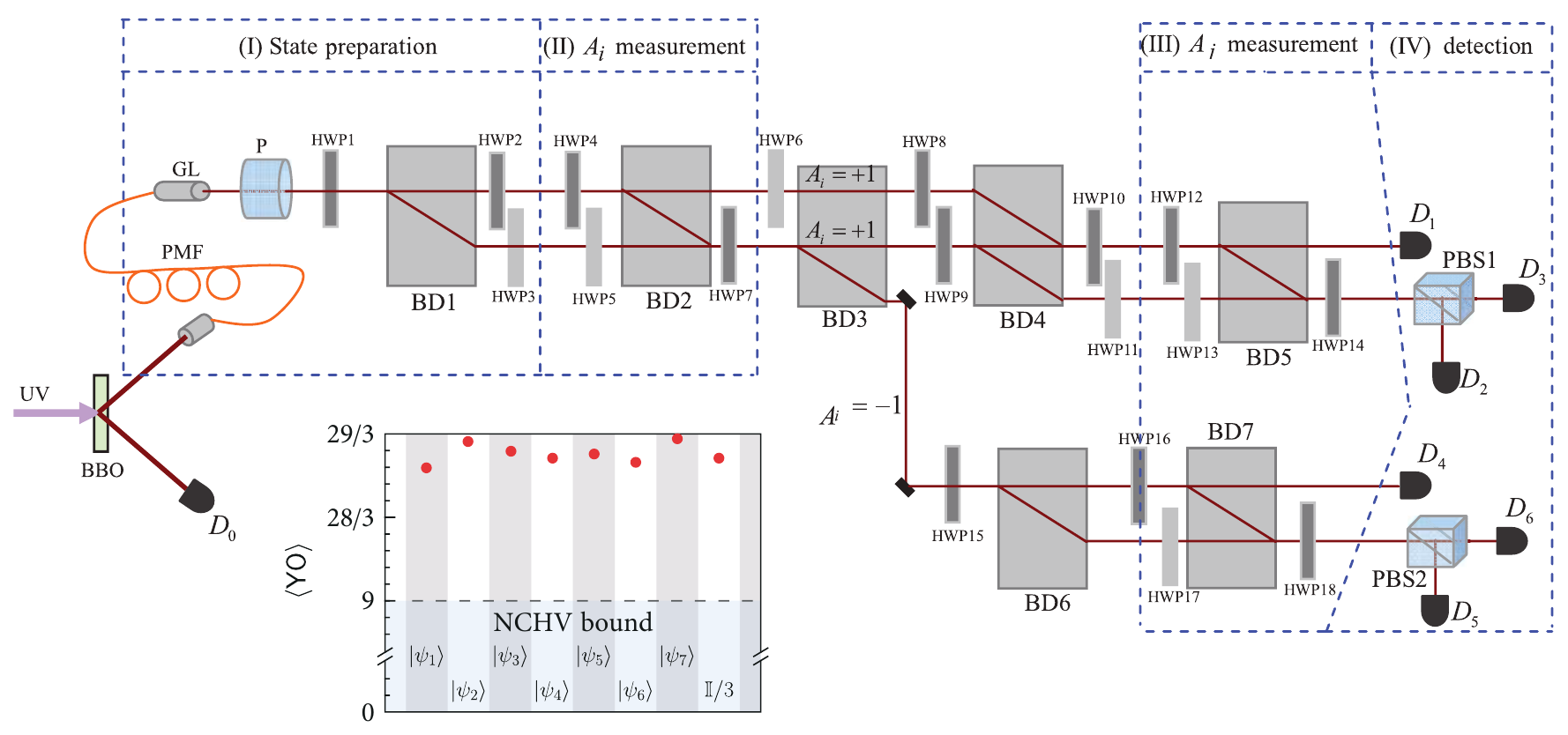}
    \caption{A ``standard'' experimental setup for testing noncontextuality inequalities containing up to two-point correlations with a photonic qutrit system. In order to extract the two-point correlation without prematurely destroying the photon, the measurement result of the first observable is registered in the path degree of freedom. Inset shows the experimental violation of the noncontextuality inequality \eqref{eq:YO_neq} for seven pure states and the maximally mixed state. Figure adapted from Reference \ignorecitefornumbering{\cite{yfhuang13}}.}
    \label{fig:setup_yfhuang13}
\end{figure*}

\subsection{Violation of noncontextuality inequality}

Having presented the above minimal example, let us now demonstrate another work by some of the same authors finished 10 years later, which can be considered a ``standard'' contextuality experiment. Here, the authors demonstrated a stringent violation of a noncontextuality inequality. As such, standard analysis of experimental errors applied and the evidence of contextuality became qualitative. \citet{yfhuang13} have followed the proposal in Reference \cite{Cabello12} to test a state-independent noncontextuality inequality based on the Yu--Oh 13-ray. The noncontextuality inequality reads:
\begin{align}
    \begin{aligned}
    \braket{\sf YO}:=& \frac{1}{2}\left(\sum_{i=1}^{4}\left\langle A_{i}\right\rangle-\sum_{i=1}^{4} \sum_{j=5}^{10} \Gamma_{i, j}\left\langle A_{i} A_{j}\right\rangle\right) \\
    &+\sum_{k=5}^{13}\left\langle A_{k}\right\rangle-\sum_{m=5}^{12} \sum_{n>m}^{13} \Gamma_{m, n}\left\langle A_{m} A_{n}\right\rangle \overset{\rm NCHV}{\leqslant} 9,
    \end{aligned}
    \label{eq:YO_neq}
\end{align}
where the definition of the observables are $A_k=\mathbb{I}_3-2\ket{a_k}\bra{a_k}$, with $\{\ket{\mathbf{a}}\}=\{\ket{\mathbf{h}}, \ket{\mathbf{y}^\pm}, \ket{\mathbf{z}}\}$ being the sequenced assemblage of the Yu--Oh 13-ray. The coefficients $\Gamma_{i,j}$ are the elements in the adjacency matrix of the Yu--Oh 13-ray's graph of exclusivity $G_{\rm YO}$ as shown in \autoref{fig:yo_vec}: $\Gamma_{i,j}=1$ if $(i,j)\in G_{\rm YO}$, $\Gamma_{i,j}=0$ if $(i,j)\not\in G_{\rm YO}$. Using quantum mechanics, it can be calculated that for any quantum state $\ket{\psi}$, $\braket{\sf YO}_\psi=29/3>9$, so the noncontextuality inequality \eqref{eq:YO_neq} is state-independently violated by any quantum state.

The noncontextuality inequality \eqref{eq:YO_neq} only involves marginal probabilities and two-point correlations. Therefore, an experimental test will need to extract these probabilities and correlations. However, when a photonic quantum is ``measured'' in the common sense, the single photon detection process will destroy the photon and prohibit the registration of two-point correlation. To address this issue, \citet{yfhuang13} registered the measurement result of the first observable on the spatial modes of single photons, so the second measurement can be implemented using the conventional photon counting technique, making the measurement of two-point correlations possible. The experimental setup, as shown in \autoref{fig:setup_yfhuang13}, is based on the beam displacer architecture \cite{Broome10}. A beam displacer is a birefringent calcite crystal with a suitably cut optical axis. When passing through the beam displacer, the vertical and horizontal polarizations of photons are separated by a fixed distance, causing the path and polarization states of the photons to become entangled.

The setup for measurement of $\braket{A_iA_j}$ comprised four main stages. Firstly, a beam displacer and two half-wave plates were employed to prepare arbitrary qutrit state. The state was encoded on the hybrid polarization--path degrees of freedom of photons, so the three computational basis were defined by $\ket{0}\leftrightarrow\ket{H}\ket{L}, \ket{1}\leftrightarrow\ket{V}\ket{L}$, and $\ket{2}\leftrightarrow\ket{V}\ket{R}$, with $\ket{H}$ and $\ket{V}$ denoting the horizontal and vertical polarizations of the photon, and $\ket{L}$ and $\ket{R}$ identifying the upper and lower paths of the photon, respectively. Secondly, a group of half-wave plates, followed by a beam displacer, implemented a basis rotation causing the $-1$-eigenstate of the operator $A_i$ to be shifted to the lower path and had a vertical polarization state. An additional beam displacer and two reflective mirrors then detached this mode from the main setup into an auxiliary setup. From now on, the evolutions in the main and auxiliary setups were made identical. Another group of half-wave plates followed by a of beam displacer then reverted the basis rotation and restored the computational basis. Thirdly, a basis rotation again shifted the $-1$-eigenstate of $A_j$ onto the computational basis. Finally, polarizing beam splitters separated the three computational basis into photodetectors, and coincidence counting was used to record the event probabilities, from which the expectation values, $\braket{A_iA_j}$, can be recovered as:
\begin{align}
    \begin{aligned}
    \left\langle A_{i} A_{j}\right\rangle=& \Pr\left(A_{i}=+1, A_{j}=+1\right)-\Pr\left(A_{i}=+1, A_{j}=-1\right) \\
    -&\Pr\left(A_{i}=-1, A_{j}=+1\right)+\Pr\left(A_{i}=-1, A_{j}=-1\right).
    \end{aligned}
\end{align}
In this way, all the necessary statistics for testing the state-independent contextuality can be observed. The experimental results were close to quantum mechanics' prediction and clearly demonstrated contextuality: even the maximally mixed state violated the noncontextuality inequality \eqref{eq:YO_neq} by over 44 standard deviations.

\subsection{Contextuality as prepare-and-measure experiments}

We see from the above example that photonic tests of contextuality relies heavily on interferometry. Arguably, the most significant obstacle for a contextuality test falls on the requirement of implementing successive measurements: if we want to test a noncontextuality inequality with $n$-point correlations, the final stage of interferometer will need to be repeated $2^{n-1}$ times. For example, to test the famous Peres--Mermin square argument of contextuality \cite{Mermin93}, an experiment will need to record three-point correlations, which means the final stage of interferometer will need to be repeated $2^{3-1}=4$ times \cite{Amselem09}. The exponential overhead of interferometry complexity poses a severe limitation on the realizations of contextuality test with even slightly complicated forms containing multi-point correlations.

\begin{figure}[tbh]
    \centering
    \includegraphics[width=.98\columnwidth]{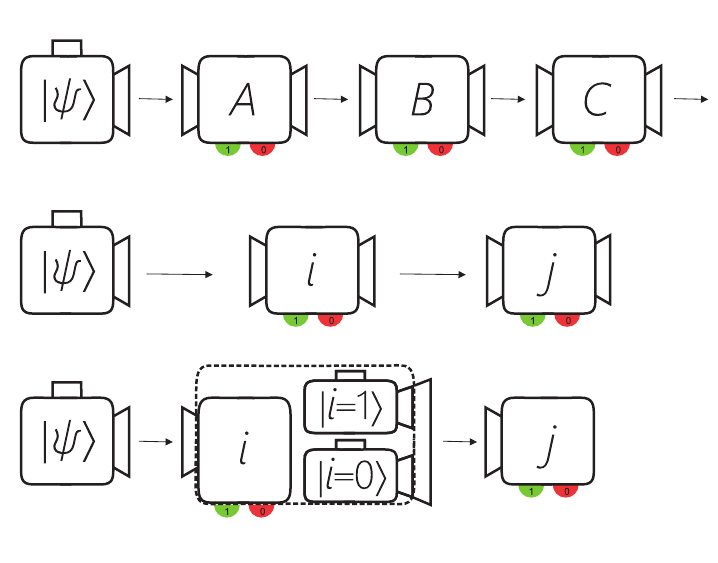}
    \caption{Simplification of contextuality experiments. Top: implementing successive measurements poses the main technical challenge on photonic contextuality experiments. Middle: By adopting the graph-theoretical approach to contextuality, the required number of sequential measurements can be reduced to one. Bottom: by assuming the L\"uders' rule, the sequential measurement can be substituted by a destructive measurement and a repreparation procedure, thus completely lifting the requirement of sequential measurements from contextuality experiments at the price of some conceptual disadvantages. Figure taken from Reference \ignorecitefornumbering{\cite{Cabello16}}.}
    \label{fig:CSW_PAM}
\end{figure}

The above problem can be partially remedied by virtue of the graph-theoretical approach to contextuality, which demonstrate that noncontextuality inequality can already be composed using up to two-point correlations (see \autoref{fig:CSW_PAM}). However, as the dimensionality of the system increases, even the architecture of two-stage interferometer in tandem will become undesirably cumbersome and introduce significant experimental imperfections. Therefore, it is worth deriving a protocol to test contextuality using only marginal probabilities instead of even two-point correlations. Because with one-stage interferometer we can only measure marginal probabilities, and the marginals in quantum theory are governed by the Born's rule which is noncontextual, this objective must be realized with some additional assumptions, probably already from quantum mechanics.

\citet{Cabello16} have proposed a method to test any form of graph-theoretical noncontextuality inequalities, with only marginal probabilities and with the assistance of the L\"uders' rule of quantum measurement. With this method, the sequential measurements in a contextuality experiment is replaced by a destructive measurement and a subsequent repreparation of a suitable quantum state; the reprepared state is calculated from the measurement outcome of the destructive measurement according to the L\"uders' rule. More specifically, the two-point correlation term in \autoref{eq:CSW_exp} shall be replaced by the product of two marginal probabilities:
\begin{align}
    \Pr(1,1|i,j)\to\Pr(1|i)\,\Pr_{\ket{i}}(1|j),
    \label{eq:reprep}
\end{align}
the subscript $\ket{i}$ indicates the corresponding probability should be measured against the nondegenerate eigenstate of the first projector to conform with the L\"uder's rule. Experimentally, the measurement--repreparation procedure in the dashed box of \autoref{fig:CSW_PAM} can either be realized with active feed-forward via electro-optic modulation or split into two different times, so the first and the second probability terms can be tested individually and even with the same setup.

\begin{figure*}[tbh]
    \centering
    \includegraphics[width=.98\textwidth]{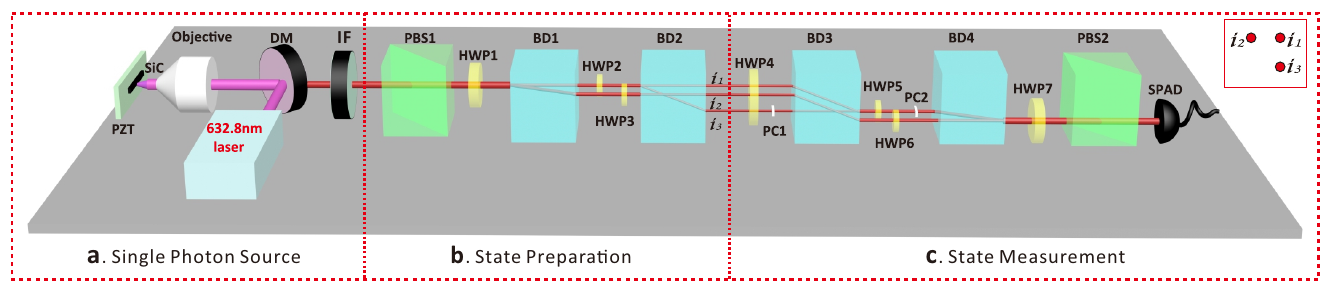} \\[12pt]
    \includegraphics[width=.98\textwidth]{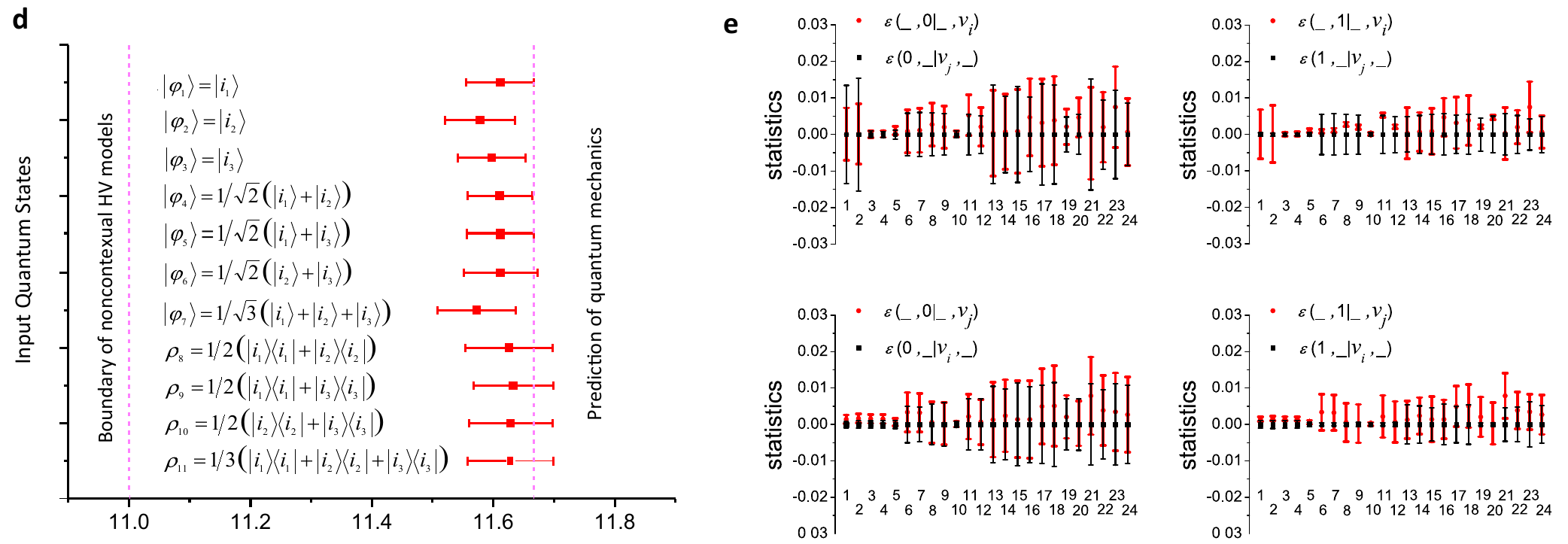}
    \caption{A photonic prepare-and-measure setup for testing graph-theoretic noncontextuality inequalities. (a)-(c) With the repreparation procedure, the two-point correlations can be calculated via \autoref{eq:CSW_exp} and \autoref{eq:reprep}. (d) Experimental results of the contextuality test. (e) Verification of the no-signaling condition. Figure taken from Reference \ignorecitefornumbering{\cite{yxiao17}}.}
    \label{fig:setup_yxiao17}
\end{figure*}

With the repreparation procedure, the complicated contextuality experiments can be reduced to the rather basic prepare-and-measure experiments. Here, we exhibit another contextuality experiment by \citet{yxiao17} based on the Yu--Oh 13-rays. The inequalities tested in this experiments and Reference \cite{yfhuang13} were similar. However, The experimental setup shown in \autoref{fig:setup_yxiao17}(a)-(c) was discernibly simpler than that in the other experiment: it only contained two stages, corresponding to the preparation and measurement procedures. The two-point correlations required in \autoref{eq:CSW_exp} were calculated with \autoref{eq:reprep} and with two different preparation and measurement procedures in the same setup, and the experimental results are given in \autoref{fig:setup_yxiao17}(d). On a more technical level, this experiment had two further differences from Reference \cite{yfhuang13}, that the photonic qutrit is entirely encoded on the path degree of freedom using one more beam displacer, and that a genuine single photon source by exciting an intrinsic defect in a silicon carbide sample \cite{jfwang18} were 
employed in the place of the heralded single photon source to eliminate the multi-photon events during the parametric process which necessitate additional compensation \cite{Lapkiewicz11,anzhang19,dkqu21}.

The photonic contextuality experiments reviewed above are inevitably based on the path degrees of freedom; next we show its role in the prepare-and-measure-based contextuality tests is not indispensable. \citet{zhliu19} realized an experiment to compare the strengths of nonlocal and contextual quantum correlations, where the contextuality test were based on an orbital angular momentum interferometry. The orbital angular momentum of photons spans an infinite-dimensional Hilbert space \cite{Allen92}, its on-demand manipulation can be achieved with a phase-only spatial light modulator \cite{Bolduc13} and its detection is feasible with the help of single-mode fibers \cite{Mair01}. The prepare-and-measure setup based on orbital angular momentum have decent scalability \cite{Bent15,zdliu17,ynsun20}. Here, the authors used this degree of freedom to encode a ququart and compared its violation of a graph-theoretic contextuality inequality with the violation of a Bell inequality by a two-qubit system; the two inequalities share the same graph of exclusivity. A gap of $\Delta\approx0.3$ was observed between the violation of \autoref{eq:CSW_exp} by a ququart system and a two-qubit system, confirming a quantum contextual correlation beyond nonlocality. Comparing with the architecture of beam displacer array, the platform based on structured light could avoid the scaling overhead for manipulating high-dimensional system; to this purpose, its accuracy of operation and detection must be further improved, and techniques like weak measurement-based wavefront sensing \cite{myang20zonal,yzheng21} may find their applications here.

Two potential loopholes come with the simplification of contextuality experiments into prepare-and-measure experiments. Firstly, the L\"uders' rule in quantum mechanics is assumed to obviate the sequential measurements. By doing so, cares must be taken to justify this additional assumption, and the experimenter is obliged to demonstrate the measurement is ideal and follows the prediction of the quantum mechanics. Secondly, the marginal probabilities themselves in a contextuality experiment must be \textit{noncontextual}. With the procedure indicated above, the reprepared state may deviate from the eigenstates of the first measurement, so the experimenter is required to explicitly test the compatibility of the two measurements. This can be accomplished by showing the marginal probability of the second measurement is not affected by the choice of the first measurement. More clearly, the signaling factors,
\begin{align}
    \varepsilon_{ij}:=\Pr(1|j)-\Pr(\_,1|i,j),    
\end{align}
can be defined over the edges of the graph of exclusivity $G$, where the underline indicates the outcome of the first measurement is irrelevant, but the measurement itself (and its associated repreparation procedure) should nevertheless be implemented. Then, an experiment with reliable compatibility relationships should show overall vanishing signaling factors: $\varepsilon_{ij}\approx0, \;\forall\, (i,j)\in E(G)$. In Reference \cite{zhliu19}, the authors reported an average signaling factor of $\abs{\bar\varepsilon}=(0.22\pm1.44)\times10^{-2}$; the details are given in \autoref{fig:setup_yxiao17}(e) .This level of signaling factor reflected close to ideal compatibilities between successive measurements and thus justified the assumptions in the simplified contextuality experiment. 

\begin{figure}[b!]
    \centering
    \includegraphics[width=.98\columnwidth]{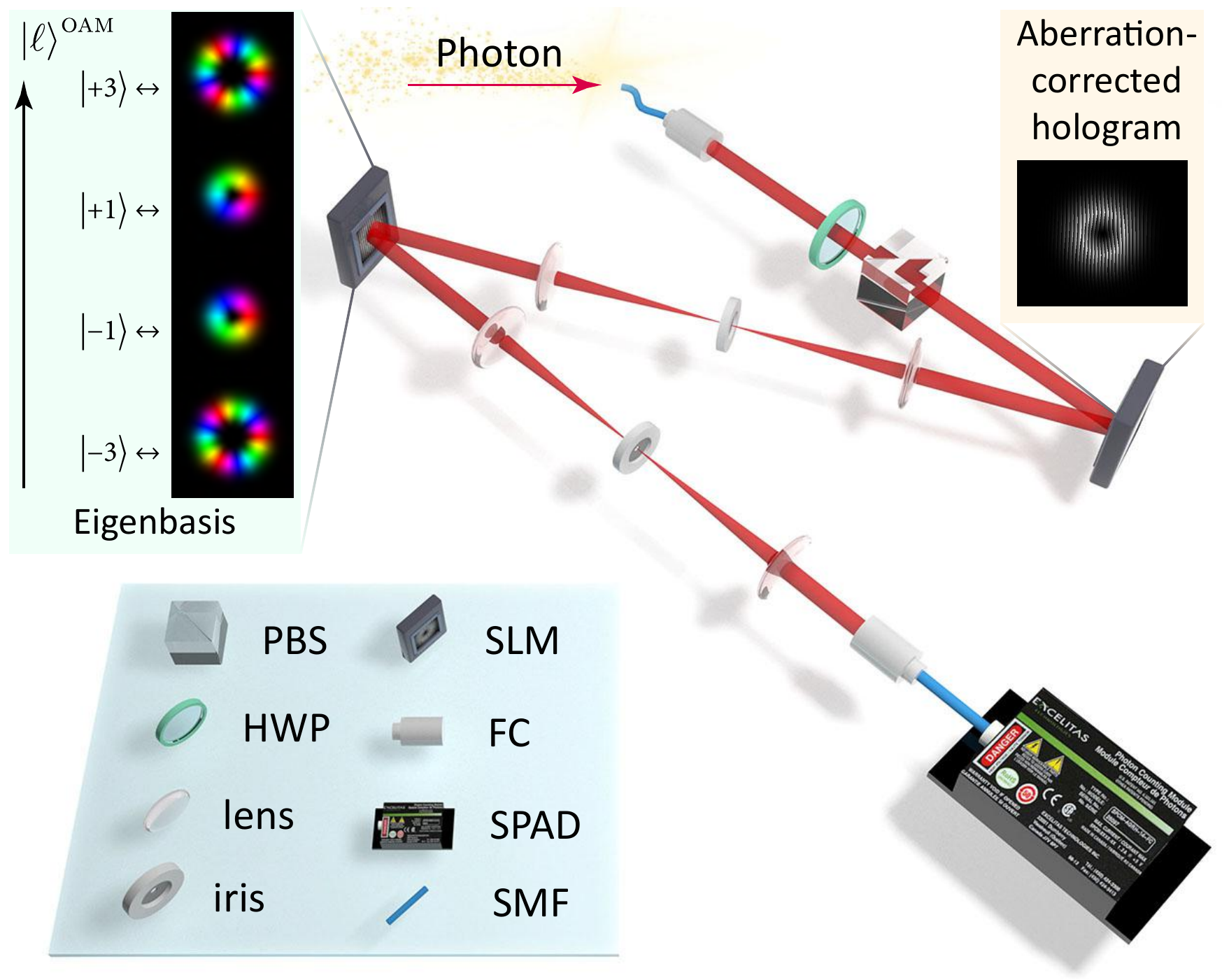}
    \caption{A high-dimensional photonic prepare-and-measure setup, where the quantum information was encoded on the orbital angular momentum degree of freedom.}
    \label{fig:setup_zhliu19}
\end{figure}

\subsection{The power of multiple degrees of freedom}

We have already shown that different photonic degrees of freedom can be utilized in contextuality experiments and have different advantages. If combining different degrees of freedom in a same experiment, they can encode more complex quantum system in which more exotic features can be observed. The idea of combining multiple degrees of freedom has become a central theme in the development of photonic quantum information processing and have broad applications. The power comes from the ability of encoding more quantum information on a single photon \cite{zyhao22,shru22}, and entangling different photonic degrees of freedom \cite{xlwang18,shru21}. 

Within the topic of contextuality, measurements corresponding to the rays in a Kochen--Specker set can give rise to ``all-versus-nothing'' contextuality paradoxes \cite{Greenberger90} (also known as quantum pseudo-telepathy \cite{Brassard03}, strong contextuality \cite{Abramsky11}, and perfect Hardy-type paradox \cite{zhliu21php} in different theoretical frameworks) which we shall subsequently discuss. Interestingly, such paradoxes were first identified in the scenario of multi-qubit Bell nonlocality and formulated with the language of Pauli observables rather than projective measurements. Experiments involving multiple degrees of freedom can effectively manipulate more qubits using the same number of photons, and thus are more suitable for observing such paradoxes.

Here, we demonstrate a concrete example \cite{Cabello01} with a four-qubit hyperentangled state $\ket{\xi}=\ket{\Psi^-}^{(12)}\otimes\ket{\Psi^-}^{(34)}$, with $\ket{\Psi^-}=(\ket{01}-\ket{10})/\sqrt{2}$ being the singlet state and the superscripts labelling the four qubits. Because $\ket{\Psi^-}$ is the common eigenstate of the Kronecker products of two identical Pauli matrices, we have: 
\begin{align}
    \begin{aligned}
    \sigma_z^{(1)}\cdot\sigma_z^{(2)}\ket{\xi}=-\ket{\xi}, &\;\; \sigma_z^{(3)}\cdot\sigma_z^{(4)}\ket{\xi}=-\ket{\xi}, \\
    \sigma_x^{(1)}\cdot\sigma_x^{(2)}\ket{\xi}=-\ket{\xi}, &\;\; \sigma_x^{(3)}\cdot\sigma_x^{(4)}\ket{\xi}=-\ket{\xi}.
    \end{aligned}
    \label{eq:avn_1}
\end{align}
Next, in reminiscence of the entanglement swapping \cite{Bennett93} procedure, we have: 
\begin{align}
    \begin{aligned}
    (\sigma_z^{(1)}\sigma_z^{(3)})\cdot\sigma_z^{(2)}\cdot\sigma_z^{(4)}\ket{\xi} &= \ket{\xi}, \\
    (\sigma_x^{(1)}\sigma_x^{(3)})\cdot\sigma_x^{(2)}\cdot\sigma_x^{(4)}\ket{\xi} &= \ket{\xi}, \\
    (\sigma_z^{(2)}\sigma_x^{(4)})\cdot\sigma_z^{(1)}\cdot\sigma_x^{(3)}\ket{\xi} &= \ket{\xi}, \\
    (\sigma_x^{(2)}\sigma_z^{(4)})\cdot\sigma_x^{(1)}\cdot\sigma_x^{(3)}\ket{\xi} &= \ket{\xi}.
    \end{aligned}
    \label{eq:avn_2}
\end{align}
Note that the operator product in the parentheses should be considered a single physical observable. Furthermore, the hyperentangled state is also an eigenstate of the following operator:
\begin{align}
    (\sigma_z^{(1)}\sigma_z^{(3)})\cdot(\sigma_x^{(1)}\sigma_x^{(3)})\cdot(\sigma_z^{(2)}\sigma_x^{(4)})\cdot(\sigma_x^{(2)}\sigma_z^{(4)})\ket{\xi} &= -\ket{\xi}.
    \label{eq:avn_3}
\end{align}
Now, all it takes to establish the all-versus-nothing contextuality is to show that a global response function cannot be defined for all these operators: by replacing the observables in \autoref{eq:avn_1} through \autoref{eq:avn_3} by their corresponding response functions, we have:
\begin{align}
    \begin{aligned}
    v(\sigma_z^{(1)})\cdot v(\sigma_z^{(2)})=-1, \;\; v(\sigma_z^{(3)})\cdot v(\sigma_z^{(4)})&=-1, \\
    v(\sigma_x^{(1)})\cdot v(\sigma_x^{(2)})=-1, \;\; v(\sigma_x^{(3)})\cdot v(\sigma_x^{(4)})&=-1, \\
    v(\sigma_z^{(1)}\sigma_z^{(3)})\cdot v(\sigma_z^{(2)})\cdot v(\sigma_z^{(4)})&=+1, \\
    v(\sigma_x^{(1)}\sigma_x^{(3)})\cdot v(\sigma_x^{(2)})\cdot v(\sigma_x^{(4)})&=+1, \\
    v(\sigma_z^{(2)}\sigma_x^{(4)})\cdot v(\sigma_z^{(1)})\cdot v(\sigma_x^{(3)})&=+1, \\
    v(\sigma_x^{(2)}\sigma_z^{(4)})\cdot v(\sigma_x^{(1)})\cdot v(\sigma_z^{(3)})&=+1, \\
    v(\sigma_z^{(1)}\sigma_z^{(3)})\cdot v(\sigma_x^{(1)}\sigma_x^{(3)})\cdot v(\sigma_z^{(2)}\sigma_x^{(4)})\cdot v(\sigma_x^{(2)}\sigma_z^{(4)}) &=-1.
    \end{aligned}
\end{align}
However, evaluating the product of all these response functions yields:
\begin{align}
    \begin{aligned}
    v(\sigma_z^{(1)})^2\; v(\sigma_z^{(2)})^2\; v(\sigma_z^{(3)})^2\; v(\sigma_z^{(4)})^2& \\
    v(\sigma_x^{(1)})^2\; v(\sigma_x^{(2)})^2\; v(\sigma_x^{(3)})^2\; v(\sigma_x^{(4)})^2& \\
    v(\sigma_z^{(1)}\sigma_z^{(3)})^2\; v(\sigma_x^{(1)}\sigma_x^{(3)})^2& \\
    \times \hspace{48pt} v(\sigma_z^{(2)}\sigma_x^{(4)})^2\; v(\sigma_x^{(2)}\sigma_z^{(4)})^2 &= -1, \\
    \midrule \midrule
    \end{aligned} \nonumber\\
    +1=-1.
\end{align}
Therefore, if we fix the measurement results of the observables in \autoref{eq:avn_1} and \autoref{eq:avn_2}, then the quantum and noncontextual hidden-variable theories will give opposite predictions on the measurement outcome of the final observable in \autoref{eq:avn_3}.

The main theoretical contribution of the above construction \cite{Cabello01} lays at that only two observers will be needed to demonstrate the paradox: pairing the qubits (1,3) and (2,4) together makes all observables local. Still, the biggest technical challenge remaining for observing such a paradox is the requirement of four-qubit hyperentanglement. It was originally suggested that two pairs of photons carrying polarization singlet states generated from spontaneous parametric down-conversion process \cite{Kwiat95} should be distributed between the two observers; however, this will require a Bell state measurement by one observer to herald the detection of another observer, and the multi-photon events will introduce systematic error even in the limit of vanishing pumping power. Fortunately, \citet{zbchen03} soon pointed out that one pair of photons will already suffice to encode two singlet states. The trick is to utilize the path degree of freedom to encode an additional singlet state by creating two possible output paths for the parametric photons via two identical down-conversion processes.

\begin{figure}[tbh]
    \centering
    \includegraphics[width=.98\columnwidth]{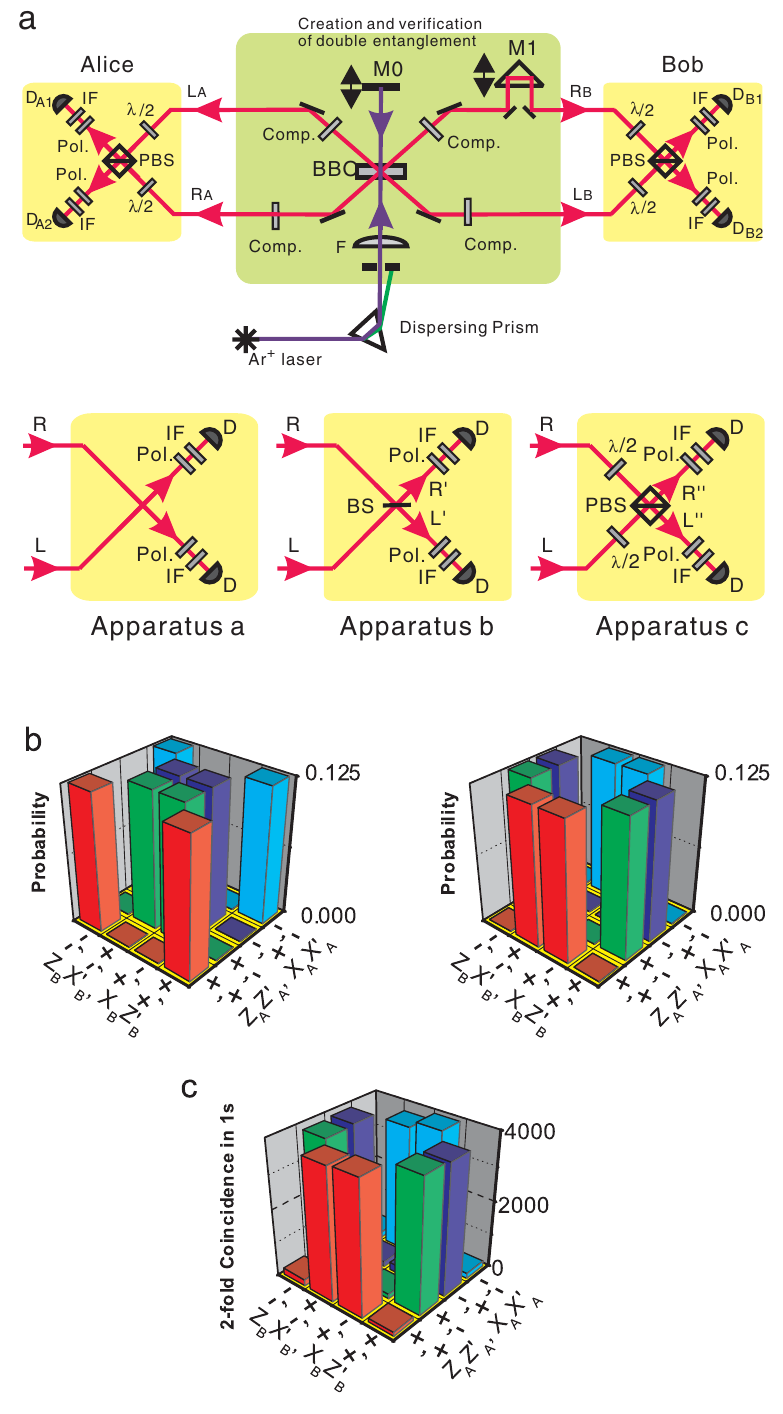}
    \caption{Observation of an all-versus-nothing contextuality. (a) Experimental setup. By pumping a nonlinear crystal twice, the two photons received by two observers became path--polarization hyperentangled. Differently apparatuses were devised to measure different path observables. (b) Predictions by noncontextual hidden-variable theory (left) and quantum theory (right) on the probabilities of events in \autoref{eq:avn_3}. (c) Experimental results and the quantum prediction are in accord. Figure taken from Reference \ignorecitefornumbering{\cite{tyang05}}.}
    \label{fig:setup_tyang05}
\end{figure}

We follow the experimental work by \citet{tyang05} to expound the idea of dual path--polarization encoding for demonstrating the all-versus-nothing contextuality. The experimental setup is shown in \autoref{fig:setup_tyang05}. A $\beta$-barium borate (nonlinear crystal) was pumped by an ultraviolet beam, where a spontaneous parametric down-conversion process can take place to generate a pair of infrared photons with entangled polarization state $\ket{\Psi^-}^{\text{pol.}}=(\ket{HV}-\ket{VH})/\sqrt{2}$. The pump beam was then reflected by a mirror to pass through the nonlinear crystal again, enabling a second down-conversion process. Subsequently, the generated photon pairs were distributed between two observers, causing the path states of the two photons to become entangled: $\ket{\Psi^-}^{\text{path}}=(\ket{LR}-\ket{RL})/\sqrt{2}$. In this manner, the four-qubit hyperentangeled state had been entangled on the two photons, where the polarization and path states of the two photons were taken as qubits (1,2) and (3,4), and the map between computational basis and physical states read $\ket{H}_{\rm pol.}\leftrightarrow\ket{0}\leftrightarrow\ket{L}_{\rm path}, \ket{V}_{\rm pol.}\leftrightarrow\ket{1}\leftrightarrow\ket{R}_{\rm path}$.

Once the map between the optical qubits and the mathematical model had been established, the observation of all-versus-nothing contextuality boiled down to certifying the constraints in \autoref{eq:avn_1}, \eqref{eq:avn_2} and testing the product of the expectations in \autoref{eq:avn_3}. The authors of Reference \cite{tyang05} realized these measurements with path interferometers and polarization analysis systems. These apparatuses are illustrated in \autoref{fig:setup_tyang05}. Concretely, the ``Apparatus a'' can measure the path observable $\sigma_z^{\rm path}$ plus an arbitrary polarization observable; here it was chosen from $\{\sigma_x^{\rm pol.}, \sigma_z^{\rm pol.}\}$. The ``Apparatus b'' can measure the path observable $\sigma_{x, \rm path}$ also plus an arbitrary polarization observable. The ``Apparatus c'' can cast a joint path--polarization measurement: the polarizing beam splitter entangled the two degrees of freedom. If without the half-wave plates before the polarizing beam splitter, a photodetection would indicate the photon comes from one of the Bell states, which are the common eigenstates of $\sigma_x^{\rm pol.}\otimes\sigma_x^{\rm path}$ and $\sigma_z^{\rm pol.}\otimes\sigma_z^{\rm path}$ \cite{NC00}. Furthermore, by adding the half-wave plates before the polarizing beam splitter, the observables $\sigma_z^{\rm pol.}\otimes\sigma_x^{\rm path}$ and $\sigma_x^{\rm pol.}\otimes\sigma_z^{\rm path}$ can also be simultaneously measured. In this way, all the probabilities constituting the observables in \autoref{eq:avn_1} through \autoref{eq:avn_3} can be registered. The experimental results for testing \autoref{eq:avn_3} are given in \autoref{fig:setup_tyang05} together with predictions from noncontextual hidden-variable theory and quantum theory. Clearly, the results were in agreement with quantum theory and demonstrated a sharp contradiction versus the axiom of noncontextuality, and no inequality was required to manifest the contradiction.

\subsection{Loophole-free tests of contextuality}

Having introduced the several contextuality tests above, we are now in a position to consider to what extent these contextuality tests serves to prove that the Nature is contextual. If a contextuality test comes with significant loopholes, the observed statistic refuting noncontextuality may actually be due to the loophole, thus the argument of contextuality will be hampered. In the light of the above argument, developing a loophole-free test of contextuality provides particularly more insights on the pertinent topic. For the photonic tests of contextuality, the imperfect single photon detection efficiency will cause some photons passing through the setup not being registered. In the most adverse scenario, all these unregistered events decrease the violation of noncontextuality inequality, so the observed phenomena could be instead resulted by the biased detection and a underlying physical law that is noncontextual \cite{Massar02}. This constitutes the so-called detection loophole. To close the detection loophole in a contextuality test, either medias of quantum information other than photons has to be chosen \cite{pfwang22}, or high-efficiency superconducting nanowire single photon detectors must be employed \cite{Shalm15,Giustina15}. If the loophole is left open, the experimenter will be obliged to accept the assumption of ``fair sampling'' indicating that the detector is plausible and does not postselect over the photons to alter the statistics that should be observed.

\begin{figure*}[t!]
    \centering
    \includegraphics[width=.8\textwidth]{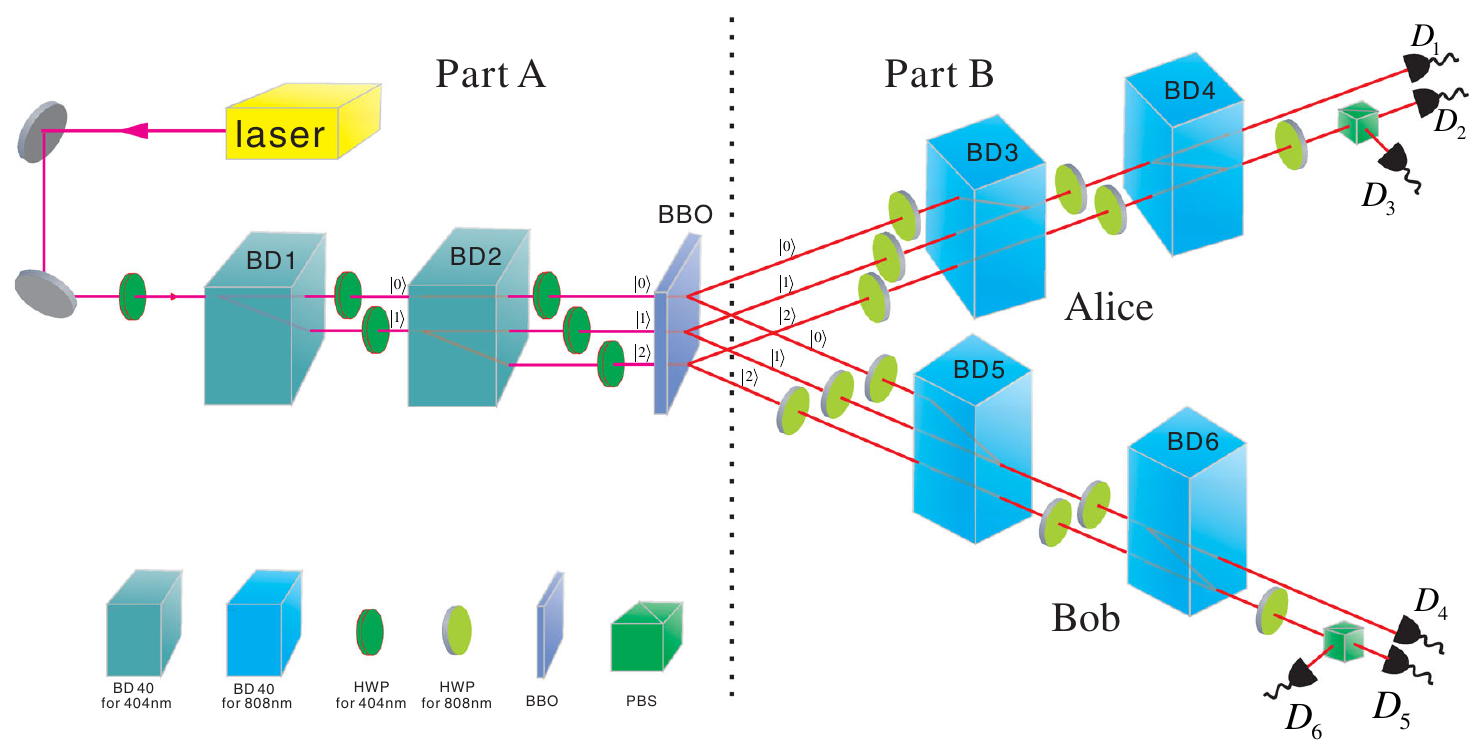}
    \caption{Experimental setup for observing a compatibility-loophole-free contextuality. Part A implemented state preparation, where maximally entangled qutrits were generated on the photonic path degrees of freedom. Part B implemented qutrit measurements, where the path states of the photons were analyzed. Figure taken from Reference \ignorecitefornumbering{\cite{xmhu16}}.}
    \label{fig:exp_xmhu16}
\end{figure*}

The other less contrived loophole in contextuality experiments originates from the imperfections of compatibility between ideally orthogonal measurements. The theory of contextuality hinges on the definition of measurement compatibility, and it was argued without perfect compatibility and infinite measurement precision, all experimental evidences supporting contextuality will be nullified \cite{Meyer99,Kent99}. The loophole can be fixed with two methods: either a generalized definition of noncontextuality that takes into account the imperfection of compatibility can be adopted \cite{Kirchmair09,Guehne10,Szangolies13}, or noncontextuality inequalities can be derived without using any sequential measurements on a single quantum system \cite{Cabello11proposal}. For the latter method, measurement on pairs of distant quantum systems facilitates the derivation of the compatibility-loophole-free noncontextuality inequality, since measurements happening in spacelike-separated regions are perfectly compatible; the no disturbance between these measurements are guaranteed by Einstein's special relativity.

\citet{xmhu16} realized an optical test of such compatibility-loophole-free contextuality with a pair of entangled qutrits. The noncontextuality inequality in this experiment, based solely upon conditional probabilities of distant measurements, reads \cite{Cabello11proposal}: 
\begin{align}
    \braket{\sf B}:=\Pr (D_{1}^{A}=1 \mid &D_{0}^{B}=1)-\Pr (T_{0}^{A}=a_{0} \mid D_{0}^{B}=1) \nonumber\\
    -&\Pr(T_{1}^{A}=a_{1} \mid D_{0}^{B}=1) \leqslant 0,
    \label{eq:clfc}
\end{align}
where the observables are defined as:
\begin{align}
    \begin{aligned}
    D_{0}^{A} &=D_{1}^{B}=|i\rangle\langle i|, \quad
    D_{1}^{A} =D_{0}^{B}=|f\rangle\langle f|, \\
    T_{0}^{A} &=a_{0}\left|a_{0}\right\rangle\left\langle a_{0}\left|+b_{0}\right| b_{0}\right\rangle\left\langle b_{0}\left|+c_{0}\right| c_{0}\right\rangle\left\langle c_{0}\right|, \\
    T_{1}^{A} &=a_{1}\left|a_{1}\right\rangle\left\langle a_{1}\left|+b_{1}\right| b_{1}\right\rangle\left\langle b_{1}\left|+c_{1}\right| c_{1}\right\rangle\left\langle c_{1}\right|, \\
    T_{0}^{B} &=a_{0}\left|b_{1}\right\rangle\left\langle b_{1}\left|+b_{0}\right| a_{1}\right\rangle\left\langle a_{1}\left|+c_{0}\right| c_{1}\right\rangle\left\langle c_{1}\right|, \\
    T_{1}^{B} &=a_{1}\left|b_{0}\right\rangle\left\langle b_{0}\left|+b_{1}\right| a_{0}\right\rangle\left\langle a_{0}\left|+c_{1}\right| c_{0}\right\rangle\left\langle c_{0}\right|,
    \end{aligned}
\end{align}
with
\begin{align}
    \begin{aligned}
    \left|a_{0}\right\rangle=(|1\rangle-|2\rangle) / \sqrt{2}, \;\; \left|a_{1}\right\rangle&=(|0\rangle-|1\rangle) / \sqrt{2}, \\
    \left|b_{0}\right\rangle=(|1\rangle+|2\rangle) / \sqrt{2}, \;\; \left|b_{1}\right\rangle&=(|0\rangle+|1\rangle) / \sqrt{2}, \\
    \left|c_{0}\right\rangle=|0\rangle,\;\;
    \left|c_{1}\right\rangle&=|2\rangle, \\
    |i\rangle=(|0\rangle+|1\rangle+|2\rangle) / \sqrt{3}, \;\;
    |f\rangle&=(|0\rangle-|1\rangle+|2\rangle) / \sqrt{3}.
    \end{aligned}
\end{align}
The inequality will be violated by two maximally entangled qutrits: choosing the quantum state as $(\ket{00}-\ket{11}+\ket{22})/\sqrt{33}$ causes $\braket{\sf B}=1/9>0$. Interestingly, the measurement setting in \autoref{eq:clfc} by one of the observers is fixed, so inequality \eqref{eq:clfc} cannot be considered as a test of nonlocality, although its form is in reminiscence of the probabilistic forms of Bell inequalities. Instead, it must be interpreted as a test of contextuality with distant quantum systems.

Experimentally, to observe the violation of \autoref{eq:clfc}, the most challenging task is the preparation of high-quality qutrit entanglement. Here, the authors realized a spontaneous parametric down-conversion array to attack this problem. Stemming from this work, the spontaneous parametric down-conversion array architecture has become the recent paradigm of high-dimensional entanglement generation \cite{xmhu20,zxliu20}. The experimental setup is depicted in \autoref{fig:exp_xmhu16}. First, using a beam displacer array, the pumping beam were evenly distributed into three paths. Then, a $\beta$-barium borate was pumped simultaneously by the three pumping beams, so a pair of parametric photons can be emitted from either of the incidental points and distributed between two observers. The propagating angle of the three possible paths of the emitted photons were identical; utilizing this parallelity, measurements of the photonic path were again implemented with beam displacer arrays by both of the receivers. Because the different wavelength of the pumping and parametric photons, the lengths of the beam displacers for preparation and measurement differed slightly. The authors reported an experimental value of $\braket{\sf B}=0.095\pm0.003$, violating inequality \eqref{eq:clfc} by 31 standard deviations to provide a strong loophole-free test of contextuality. 

\section{Advances and applications}
\label{sec:application}

In this section, we switch our focus on the applications of contextuality. As already mentioned briefly in the introductory paragraphs, contextuality has been found to have broad application in the general quantum information science, including quantum cryptography \cite{Svozil06,Cabello11}, quantum communication \cite{Spekkens09,Saha19}, randomness expansion \cite{Abbott12,Um20}, self-testing \cite{Bharti19} and dimension witnessing \cite{Guehne14,Ray21}. Here, we only discuss two advances in details here, that how contextuality is related to universal quantum computation, and how contextuality activates nonlocality so the two resource for quantum computation and quantum communication can be inter-converted.

\subsection{Towards universal quantum computation}

Many approaches to quantum computation has been proposed in pursue of computing power beyond the classical supercomputers. However, the computing power of current noisy intermediate-scale quantum \cite{Preskill18} circuit is severely limited by the omnipresent noise that causes the quality of computing to deteriorate. 
If the computation accuracy falls below some certain level, the quantum advantage over classical computers will vanish. Fortunately, the situation can be radically overturned if the noise of the quantum circuit can be suppressed below a critical level \cite{Shor95}. In this case, a properly designed error-correction code is suffice to asymptotically suppress any residual noise. 

The noise in a quantum circuit may occur both in the state preparation stage or during the transformation induced by quantum gates; if the transformation process is made noiseless, the errors from state preparation will not propagate, and the quantum computation will become accurate. Although it is not practical to make all transformations noiseless, it is possible for only some subsets of transformations. For example, the braiding of non-abelian anyons effectively implements noiseless Clifford gates on the encoded quantum information \cite{Nayak08}. With these ideal Cifford gates, only one ideal non-Clifford gate is required to achieve universal, fault-tolerant quantum computation \cite{Gottesman98}. As a proxy to obtain such an ideal non-Clifford gate, \citet{Bravyi05} proposed a subroutine now known as magic state distillation. The subroutine is based on the observation that a non-Clifford gate can be emulated by a controlled Clifford gate plus an ancillary quantum system, starting from a ``magic state'' away from the eigenstates of all Clifford operators and subjecting to a postselection. Magic state distillation allows the preparation of asymptotically ideal magic states with noisy magic states and Clifford quantum gates. However, magic state distillation also shows a threshold behavior: only when the noisy magic states have enough fidelity with ideal state does the subroutine increase its fidelity. A question naturally arise: what intrinsic property of a quantum system decides its usage in magic state distillation?

\citet{Howard14} showed that the decisive property of a quantum state for quantum computation is but contextuality by proving that all quantum states useful in magic state distillation violate a noncontextuality inequality constructed with Clifford operators. For a single quantum system, violation of such a magic noncontextuality inequality is equivalent to manifestation of negativity in the discrete Wigner function \cite{Veitch14}. Here, we explicitly give the form of the magic noncontextuality inequality for a qutrit system as an example. We start from the definion of the Weyl--Heisenberg displacement operators:
\begin{align}
	D_{x, z}=\omega^{2^{-1}xz}\tau^x\sigma^z,~~~ \{x, z\}\in\{0, 1, 2\}.
\end{align}
\add{Here, the operators $\tau$ and $\sigma$ are the three-dimensional shift and clock matrices defined as:
\begin{align}
		\tau &=\begin{pmatrix}
			0 & 0 & 1 \\
			1 & 0 & 0 \\
			0 & 1 & 0
		\end{pmatrix},\quad \text{and}\; \sigma = \begin{pmatrix}
		1 & 0 & 0 \\
		0 & \omega & 0 \\
		0 & 0 & \omega^2
	\end{pmatrix};
\end{align}
they are analogous to the Pauli matrices $\sigma_x$ and $\sigma_z$ in the two-dimensional case.} These operators has a spectrum of $\{1, \omega, \omega^2\}$, with \add{$\omega=e^{2\pi i/3}$}. We then denote the list of displacement operators $\mathbf{D}=\{D_{0, 1}, D_{1, 0}, D_{1, 1}, D_{1, 2}\}$, whose eigenstates span a complete set of mutually unbiased bases, and a set of magic contextuality witnessing operators:
\begin{equation}
	A^\mathbf{r}=\mathbb{I}_3-\sum_{j=1}^4\Pi_j^{r_j}, 
\end{equation}
in which $\Pi_j^{r_j}$ is the projector of the eigenstate $\omega^{r_j}$ of the $j$-th element in $\mathbf{D}$, and the definition of the vector $\mathbf{r}$ reads: $\mathbf{r}=x\mathbf{a}+z\mathbf{b}$ with $\mathbf{a}=\{1, 0, 1, 2\}, \mathbf{b}=-\{0, 1, 1, 1\}$ and $\{x, z\}\in\{0, 1, 2\}$. Using the above notations, the magic noncontextuality inequality can be stated as:
\begin{align}
	\braket{\mathcal{M}} &:= \max_{\bf r} \mathrm{Tr}\left[A^\mathbf{r} \rho \right] \overset{\mathrm{NCHV}}{\leq} 0.
	\label{eq:resource}
\end{align}
The inequality \eqref{eq:resource} can be violated up to the inverse of golden ratio, $(\sqrt{5}-1)/2$, by the magic states. It quantifies the efficacy of a quantum state for implementing ancilla-based non-Clifford gates: with the perfect magic state maximally violating inequality \eqref{eq:resource}, a noiseless non-Clifford gate can be executed and no distillation process is required; if the inequality is non-maximally violated, then some rounds of magic state distillation are in order for suppressing the noise of the non-Clifford gates below threshold \cite{Bravyi05}; if the inequality is not violated then the noise level of the distillation process becomes classically simulable, so the quantum advantage vanishes.

\begin{figure*}[t]
    \centering
    \includegraphics[width=1.00\textwidth]{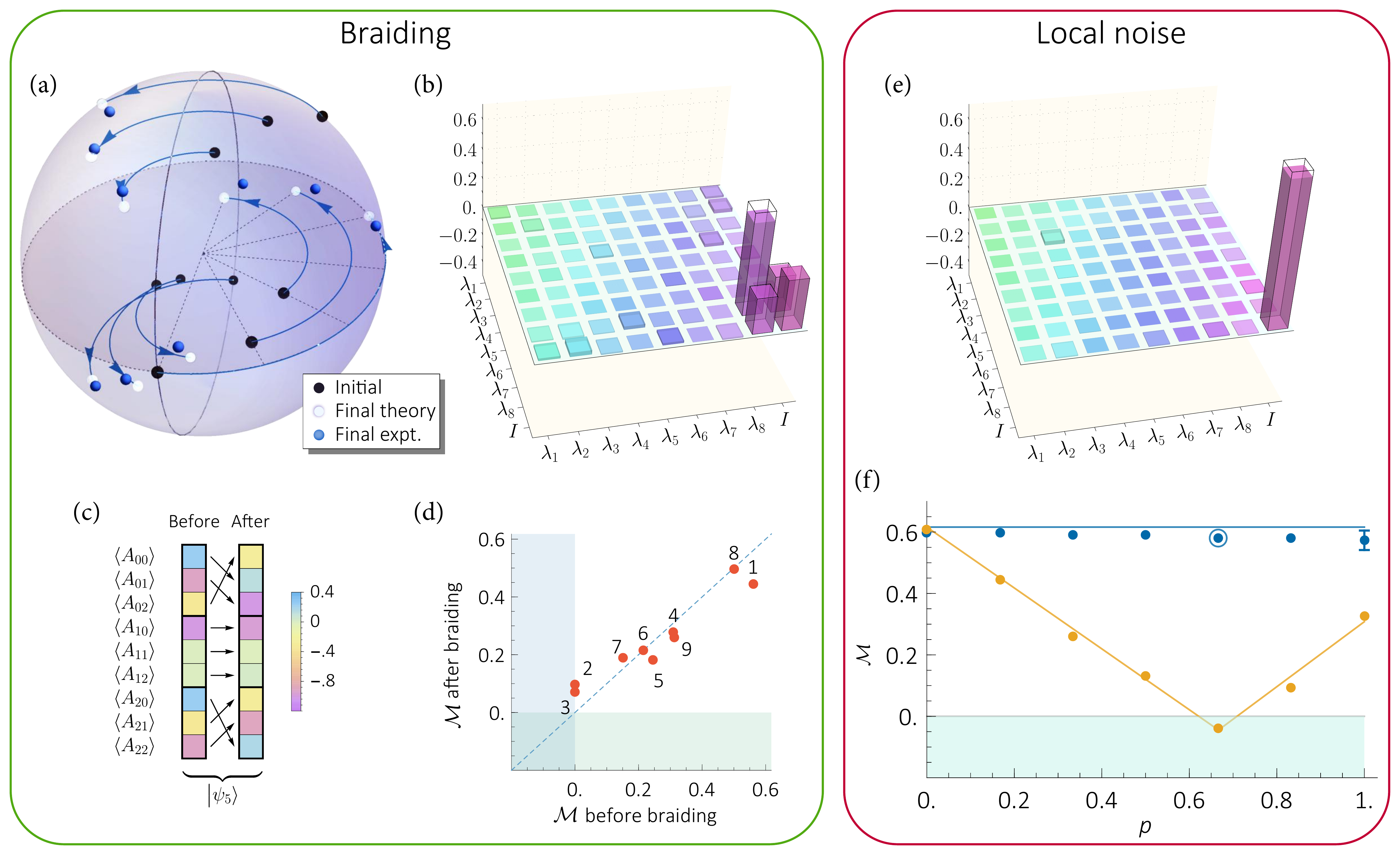}
    \caption{Behaviors of topological contextuality under braiding and local noise. (a) A Bloch sphere illustrating the effect of $\mathbb{Z}_3$-parafermion braiding on the encoded qutrit. A geometric phase of $2\pi/3$ is obtained between one and the other two computational basis. (b) The real parts of the theoretical (edge) and experimental (filling) braiding process matrices. (c) Permutation of the magic contextuality witnesses. (d) The violations of magic noncontextuality inequality \eqref{eq:resource} were almost invariant for the sample states. (e) The real parts of the theoretical (edge) and experimental (filling) local noise process matrices. (f) The dynamics of magic contextuality under hopping noise for a topologically-protected qutrit (blue) and under flip noise for a trivial (not protected) qutrit (orange). Figure adapted from Reference \ignorecitefornumbering{\cite{zhliu21pf}}.}
    \label{fig:zhliu21pfc}
\end{figure*}

The roles of contextuality in quantum computation based on non-abelian anyons appear in twofold. Firstly, the resource of magic measured by the violation of inequality \eqref{eq:resource} is invariant under Clifford gates \cite{Veitch14}. Therefore, the usefulness of a specific quantum state for magic state distillation is unaffected by braiding operations. By this observation, high-fidelity non-Clifford gates induced by magic state distillation can be executed at any point of a compiled quantum circuit. Secondly, as any local noise emerging during a topological computation is exponentially suppressed by the excitation gap \cite{Kitaev03}, the resource of magic should also be protected by the system topology. By this observation, having access to arbitrary braiding operations and infinite supply of non-perfect magic states already enables fault-tolerant universal quantum computation. Nonetheless, currently the realization of anyons in physical systems is still in its infancy \cite{Bartolomei20,Nakamura20} and faces many technical challenges; furthermore, observation of their non-abelian statistics is still intractable. 

Taking an alternative approach, \citet{zhliu21pf} studied the non-abelian statistics of anyons and its application in quantum computation with a designated photonic quantum simulator \cite{Aspuru-Guzik12,Nori14}. The authors studied a one-dimensional chain of $\mathbb{Z}_3$-parafermions (a type of non-abelian anyons) by mapping the parafermionic chain to the state space of spin-1 bosons through the Fradkin--Kadanoff transformation \cite{FK80}. With interaction parameters chosen as appropriate, a pair of parafermion-edge zero modes will emerge at the end of the chain, on which a topologically-protected qutrit immune to any local noise can be encoded. 

To elucidate the roles of contextuality in topological quantum computation, Reference \cite{zhliu21pf} directly tested the dynamics of magic contextuality under braiding evolution and local noise. By tuning the interacting Hamiltonian of the parafermionic chain, the parafermion-edge zero modes can be driven through the chain to induce the braiding evolution and topologically-protected gate operations. Experimentally, the modulation of system Hamiltonian $\cal H$ was realized by beam displacer arrays and polarizing beam splitters. These dissipative elements caused discrete imaginary-time evolution $e^{-\mathcal{H}t},\; t\to+\infty$ to project the encoded wavefunction on the ground state of $\cal H$. The geometric phase inducing the braiding evolution was preserved despite the discreteness of the evolution \cite{Bargmann64}; this correspondence was originally exploited in Reference \cite{jsxu16} to optically simulate the geometric phases induced by Majorana zero modes' braiding. On the other hand, the local noise was introduced by repopulating the optical wavefunction according to the form of the anyonic noise, translated again from the Fradkin--Kadanoff transformation, and subsequently dissipating the modes corresponding to the excited states. We note that the method of ground state generation here effectively implements a non-Hermitian Hamiltonian $\tilde{\mathcal{H}} =-i \cal{H}$, so it can also find applications in the investigation of non-Hermitian physics in, e.g., the (anti-) parity--time symmetric systems \cite{jstang16,ytwang20,ymeng20,qli19}. Besides, it is still effective even for an unknown $\cal H$ given as a controlled oracle and $t\not\to\infty$; in this setting, the process is otherwise known as algorithm cooling \cite{jsxu14}. 

To demonstrate the resource of magic's invariance under Clifford operations, the authors implemented an analogous braiding of photonic modes with the discrete imaginary-time evolution gates to generate a phase gate, so one of the computational basis of the qutrit acquired an additional $2\pi/3$ geometric phase. The effect of the analogous gate operation can be best seen in \autoref{fig:zhliu21pfc}(a), where the sample states orbited the Bloch sphere by roughly $120^\circ$ after braiding. The quantum process tomography \cite{Chaung97,Obrien04} of the braiding evolution, as shown in \autoref{fig:zhliu21pfc}(b), also helped to confirm this effect, showing a process fidelify of 93.4\% comparing to the theoretical value. Next, the left hand side values of the magic noncontextuality inequality were directly measured for the nine sample states before and after braiding. The effect of braiding on the contextuality observations can be most intuitively seen in \autoref{fig:zhliu21pfc}(c), where some of the observation's expectations were permuted. As the final measure of magic contextuality is defined over the maximum of the contextuality observations and is not affected by the permutation among different observations, the resource of magic (see \autoref{fig:zhliu21pfc}(d)) was almost invariant, apart from some experimental imperfections, before and after the braiding process.

Regarding the noise resilience of magic contextuality against local noise, the authors also exploited quantum process tomography to characterize the effect of a local hopping noise in a parafermion system. After casting the noise-induced incoherent error with some certain probabilities, the system was projected back into the ground state subspace. The process matrix as shown in \autoref{fig:zhliu21pfc}(e) was almost an identity matrix. Next, the magic contextuality value of a quantum state in the proximity of a magic state was measured before and after the disturbance--dissipation process. As the quantum state was almost not affected by the local noise apart from some probability amplitude damping, so was its degree of contextuality; this is true even in the limit of large error probability: as can be seen in \autoref{fig:zhliu21pfc}(f), a value of $\mathcal{M}=0.580\pm0.013$ was observed even at error probability $p=99\%$. For comparison, the effect of noise on a non-topologically-protected trivial qutrit was also investigated; in this scenario, the dissipation process is not implemented, and the noise quickly destroyed the resource of magic contextuality.

\subsection{Activation of nonlocality from contextuality}

Contextuality and nonlocality in quantum mechanics have deep-rooted connections: both of them are indispensable resources for quantum information tasks; behaviors of nonlocality can be seen as special forms of contextuality \cite{Cabello21bks} and every noncontextuality inequality can be converted to a Bell inequality \cite{Cabello21converting}; and the ability of a single quantum system manifesting the most elementary forms of contextuality \cite{KCBS08} and nonlocality \cite{CHSH69} has a trade-off relationship \cite{Kurzynski14,xzhan16}. The final observation puts forward a fundamental question---whether single-particle contextuality and two-party nonlocality can coexist. The answer is trivially positive if we choose to observe state-independent contextuality on one of the two particles constituting high-dimensional entangled states. However, it becomes highly intriguing if the observed contextuality promotes a quantum correlation that initially cannot violate a Bell inequality to a nonlocal correlation. To this end, experimental proposals based on two pairs of hyperentangled qubits \cite{Cabello10} and a pair of maximally entangled qutrits \cite{Cabello12} have been put forward.

\begin{figure}[tbh]
    \centering
    \includegraphics[width=1.00\columnwidth]{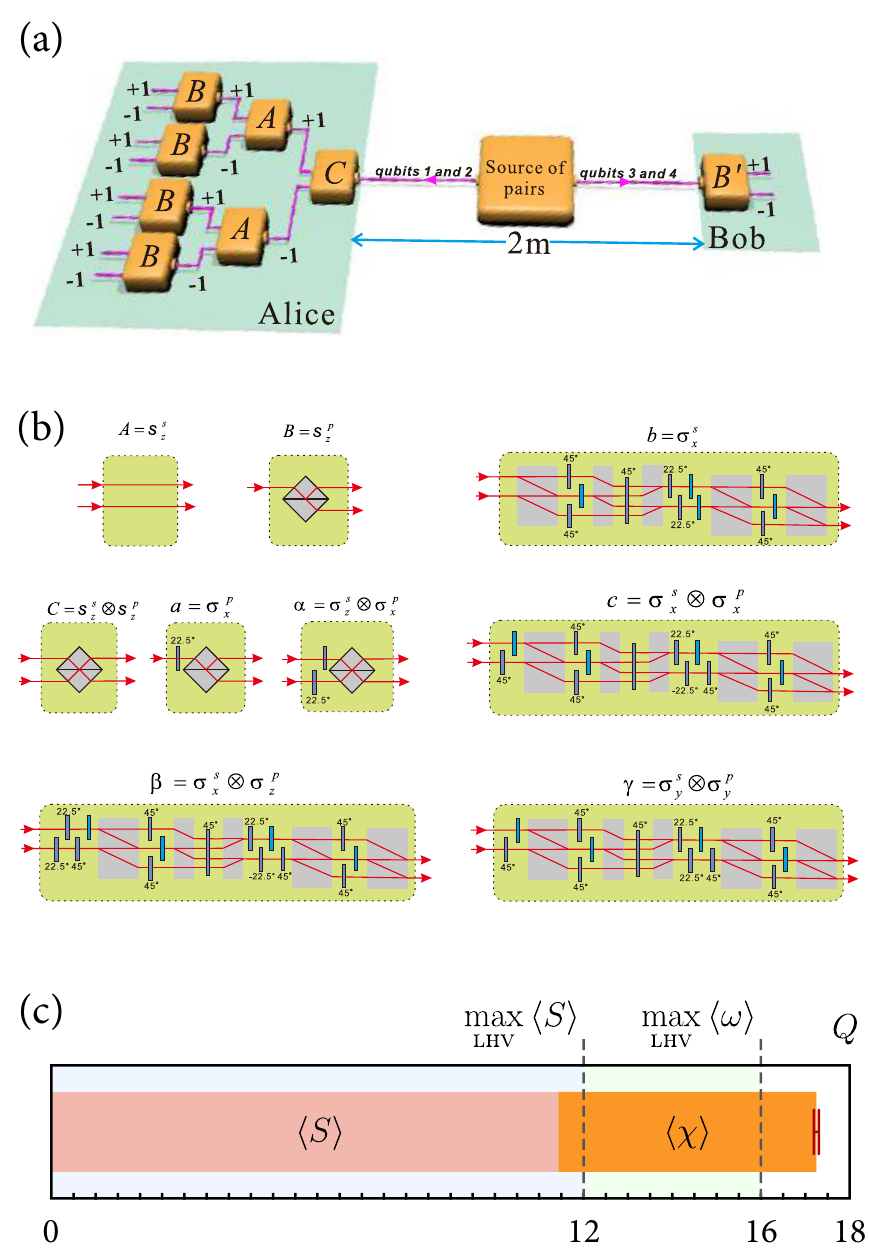}
    \caption{Setup for observation of nonlocality activated by local contextuality. (a) Schematic illustration of the experiment. Alice and Bob shares two pairs of Bell states. Alice implements a contextuality test on her qubits and also checks the correlations of her observables with Bob's. (b) Devices for measuring the nine observables in the contextuality test. (c) Experimental result demonstrating the activation of nonlocality from local contextuality. Figure adapted from Reference \ignorecitefornumbering{\cite{bhliu16}}.}
    \label{fig:exp_bhliu16}
\end{figure}

\citet{bhliu16} and \citet{xmhu18} realized the proposals in Reference \cite{Cabello10,Cabello12} with a path-polarization hyperentangled and a high-dimensional entangled photonic state. This duo of experiments shared a similar conceptual  basis; here we choose Reference \cite{bhliu16} to exemplify the underlying idea. The objective was to test a local hidden-variable inequality whose violation signifies nonlocality:
\begin{align}
    \braket{\omega}:=\braket{\chi}+\braket{S}\overset{\rm LHV}{\leqslant}16,
    \label{eq:nlqc}
\end{align}
where the two quantities $\chi$ and $S$ measure the degree of contextuality and the strength of bipartite correlations. Explicitly, $\chi$ is defined as:
\begin{align}
    \begin{aligned}
    \langle\chi\rangle=\langle C A B\rangle &+\langle c b a\rangle+\langle\beta \gamma \alpha\rangle+\langle\alpha A a\rangle+\langle\beta b B\rangle-\langle c \gamma C\rangle, 
    \end{aligned}
\end{align}
which is a sum of single particle sequential measurements' expectations. On the other hand, $S$ is defined as:
\begin{align}
    \langle S\rangle=&\left|\left\langle A A^{\prime}\right\rangle_{C A B}\right|+\left|\left\langle B B^{\prime}\right\rangle_{C A B}\right|+\left|\left\langle b b^{\prime}\right\rangle_{c b a}\right| \nonumber\\
    &+\left|\left\langle a a^{\prime}\right\rangle_{c b a}\right|+\left|\left\langle\gamma \gamma^{\prime}\right\rangle_{\beta \gamma \alpha}\right|+\left|\left\langle\alpha \alpha^{\prime}\right\rangle_{\beta \gamma \alpha}\right| \nonumber\\
    &+\left|\left\langle A A^{\prime}\right\rangle_{\alpha A a}\right|+\left|\left\langle a a^{\prime}\right\rangle_{\alpha A a}\right|+\left|\left\langle b b^{\prime}\right\rangle_{\beta b B}\right| \nonumber\\
    &+\left|\left\langle B B^{\prime}\right\rangle_{\beta b B}\right|+\left|\left\langle\gamma \gamma^{\prime}\right\rangle_{c \gamma C}\right|+\left|\left\langle C C^{\prime}\right\rangle_{c \gamma C}\right|,
\end{align}
where the prime superscript indicates the operator acts on a different quantum system, and the subscripts specify the measurement contexts of the un-primed observables. The overall experimental schematic is shown in \autoref{fig:exp_bhliu16}(a). Crucially, it is not possible to compose a meaningful Bell inequality with only quantities in $S$, since all the correlation terms have a plus sign and the local bound 12 already saturates its algebraic maximum. 

The situation becomes different only when the effect of contextuality is taken into consideration:  \autoref{eq:nlqc} is a genuine Bell inequality that can be violated to 18, again its algebraic maximum, using a pair of singlet states $\ket{\xi}=\ket{\Psi^-}^{\rm pol.}\otimes\ket{\Psi^-}^{\rm path}$. To test the inequality \eqref{eq:nlqc} experimentally, a photonic hyperentanglement source is employed in the first place; the similar technique was already discussed in \autoref{sec:experiment}. In order to achieve the maximal quantum value, the definition of the observables should be chosen as those appearing in the Peres--Mermin square  \cite{Mermin93}:
\begin{align}
    \begin{aligned}
    &A=\sigma_{z}^{\rm path.}, \; B=\sigma_{z}^{\rm pol.}, \; C=\sigma_{z}^{\rm pol.} \otimes \sigma_{z}^{\rm path}, \\
    &a=\sigma_{x}^{\rm pol.}, \; b=\sigma_{x}^{\rm path}, \; c=\sigma_{x}^{\rm pol.} \otimes \sigma_{x}^{\rm path}, \\
    &\alpha=\sigma_{x}^{\rm pol.} \otimes \sigma_{z}^{\rm path}, \; \beta=\sigma_{z}^{\rm pol.} \otimes \sigma_{x}^{\rm path}, \; \gamma=\sigma_{y}^{\rm pol.} \otimes \sigma_{y}^{\rm path},
    \end{aligned}
\end{align}
the choice of the primed obsevables should be identical to that of the un-primed obserables. An experimenter will need to implement sequence measurements to extract the six correlations appeared in $\omega$. The measurement apparatuses are shown in \autoref{fig:exp_bhliu16}(b), in short, each of these apparatuses moved the $\pm1$-eigenstate of the interested observable to the upper and lower path, respectively. 

By cascading these apparatuses, the expectation values of the correlations appeared in $\braket{\chi}$ can finally be determined. In the setting of contextuality experiment, the cascading technique was first demonstrated by \citet{Amselem09} and subsequently adopted in various works \cite{Amselem12,dAmbrosio13}, and in References \cite{dkqu20,dkqu21} in a projective manner. Here, the authors imported this technique into the beam displacer array architecture where the phases between different optical paths are essentially free of drift; thanks to the self-stabilized interferometer, a very high value of contextuality witness $\braket{\chi}=5.817\pm0.011$ was reported. On the other hand, the nonlocal correlations in $\braket{S}=11.430\pm0.016$ were determined by joint path-polarization measurements of the other two qubits at the remote site, and comparison between the outcomes of the corresponding observables. Combining these two results, a value of $\braket{\omega}=17.247\pm0.019$ was obtained, rejecting the prediction by local hidden-variable models with strong confidence. These experimental results are depicted in \autoref{fig:exp_bhliu16}(c). Besides clarifying how to produce nonlocality from local contextuality, this work also advanced the tests of Peres--Mermin square-type contextuality, in the sense that the observed phenomena subject to neither an equivalent classical description \cite{Frustaglia16} nor a decrease of contextuality visibility caused by imperfect photon-number-resolving detection \cite{bhliu09}.

\section{Discussion and outlook}
\label{sec:discussion}

We have reviewed some theoretical studies and experimental tests of contextuality, as well as some recent results demonstrating its applications, that have taken place in USTC in the last 20 years. We hope such a review is timely and relevant, since it has become clear only recently that contextuality is an indispensable resource for quantum computation \cite{Howard14} and the effect of contextuality is experimentally testable \cite{Cabello08}. Therefore, although the record of the simplest proofs of contextuality has been sealed \cite{Cabello96,YO12,Cabello15,zpxu20peres}, many theoretical topics regarding contextuality are still left unaddressed. Here, we point out two of such questions. Firstly, the examples of contextual correlations that scale linearly with the dimension of quantum system are sparse \cite{Cabello10macro,Vidick11}. This observation is in contrast to both the theoretical limit \cite{Amaral15} and the situation in the study of nonlocality, where violations of Bell inequality that scale exponentially with the qubit number have been long recognized \cite{Mermin90,Ardehali92,Belinskii93}. Although these Bell inequalities themselves can be trivially interpreted as noncontextuality inequalities, it is more intriguing to search for other noncontextuality inequalities that may have still larger quantum--classical separation. Secondly, within the framework of all-versus-nothing contextuality, currently known examples uses at least four argument clauses for demonstrating state-dependent contextuality and five for state-independent contextuality; it is thus worth exploring if this number could be further reduced. The discovery of such stronger forms of contextuality may have implications in further accelerating quantum computation.

The future directions of contextuality from the experimental side are also diverse. From a fundamental perspective, contextuality as a general phenomenon may go beyond the framework of quantum mechanics and hidden-variable models. For example, References \cite{Cavalcanti18} and \cite{Pearl21} have shown a faithful classical causal model satisfying non-disturbance between conditional probabilities always results in noncontextual correlations. Experiments in this direction may act as a proxy for detecting the compatibility of quantum mechanics and various general probabilistic theories. From the view of quantum information science, contextuality enables novel applications like self-testing of a single quantum system \cite{Bharti19}. Investigation of such properties \cite{xmhu22} promotes our ability to certify quantum apparatus with minimal assumptions. Regarding the aspects of quantum computation, quantum simulation of subroutines for quantum computation in different physical systems \cite{jsxu18,cliu19,hlhuang21} may supply additional insights, and the realization of contextuality-empowered algorithms that works with noisy intermediate-scale quantum devices, like reported in Reference \cite{Kirby21}, is also highly relevant. Finally, novel experimental systems like solid-state color centers \cite{Widmann15,jfwang20,jfwang21} may found their role in contextuality experiments. With an intrinsic quantum memory, a support of exotic operations, and the possibility of hosting macro-scale quantum entanglement, these systems may serve to investigate different forms of contextuality \cite{zpxu16,Leifer20,zpxu21rank}.

As the topic of contextuality is too broad for even a dedicated book, and it is impossible to contain all the relevant results achieved in USTC in a short review article, we are compelled to choose over different works and have striven to enlarge the scope of this review. The confluence of many exciting advances reflects both the significant role of contextuality in the quantum foundation, and the profound accumulation of research power that the university possesses. We believe the study of contextuality will boost the development of quantum technology and finally benefits the human society, and we hope that in the next twenty years USTC will proceed to spearhead in the study of contextuality and quantum information science.

\begin{acknowledgments}

We thank Ad\'an\ Cabello, Jing-Ling\ Chen, Yong-Jian\ Han, Xiao-Min\ Hu, Hui-Xian\ Meng, Jiannis\ K.\ Pachos, Hong-Yi\ Su, Ya\ Xiao, Zhen-Peng\ Xu, Xiang-Jun\ Ye and Jie\ Zhou for insightful discussions over the topics of this review.

This work was supported by
Innovation Program for Quantum Science and Technology (Grants No.\,2021ZD0301400)
%
%
the National Natural Science Foundation of China (Grants
No.\,61725504,
No.\,11774335, No.\,11821404,
and No.\,U19A2075),
the Fundamental Research Funds for the Central Universities (Grant No.\,WK2030380017 and No.\,WK2030000056), 
Anhui Initiative in Quantum Information Technologies (Grants No.\,AHY020100, and No.\,AHY060300).

\end{acknowledgments}

\vspace{20pt}
\noindent \textbf{\textsf{Conflict of interest}}
\\[12pt]
The authors declare no conflict of interest.

\vspace{20pt}


\bibliographystyle{apsrev4-2}
\bibliography{justrev}


\clearpage

\end{document}